\documentclass[%
 reprint,
 amsmath,amssymb,
 aps,
]{revtex4-2}

\usepackage{color,soul}
\usepackage{multirow} 

\usepackage{float}
\usepackage{graphicx}
\usepackage{dcolumn}
\usepackage{bm}
\usepackage[utf8]{inputenc}
\usepackage{array}
\usepackage{tabularx}
\usepackage{amsmath}
\usepackage[dvipsnames]{xcolor}
\usepackage{natbib}
\usepackage{graphicx}

\begin{document}
\newcolumntype{C}[1]{>{\centering\let\newline\\\arraybackslash\hspace{0pt}}m{#1}}
\title{Power deposition studies for standard and crystal-assisted heavy ion collimation in the CERN Large Hadron Collider}
\thanks{Research supported by the HL-LHC project.}%

\author{J.B. Potoine$^{1,2}$, R. Bruce$^2$, R. Cai$^{2,3}$, F.~Cerutti$^2$ M. D'Andrea$^2$, L. Esposito$^2$, P.D. Hermes$^2$, A. Lechner$^2$,\\D. Mirarchi$^2$, S. Redaelli$^2$, V. Rodin$^2$, F.~Salvat~Pujo$^2$, P.~Schoofs$^2$, A. Waets$^2$, F. Wrobel$^1$}

\affiliation{$^1$ Universit\'e de Montpellier, 34090 Montpellier, France\\
$^2$ European Organization for Nuclear Research (CERN),\\ Esplanade des Particles 1, 1211 Geneva, Switzerland\\
$^3$ \'Ecole Polytechnique F\'ed\'erale de Lausanne (EPFL), CH-1015 Lausanne, Switzerland }

\begin{abstract}
The LHC heavy-ion program with $^{208}$Pb$^{82+}$ beams will benefit from a significant increase of the beam intensity when entering its High-Luminosity era in Run~3 (2023). The stored energy is expected to surpass 20~MJ per beam. The LHC is equipped with a betatron collimation system, which intercepts the transverse beam halo and protects sensitive equipment such as superconducting magnets against beam losses. However, nuclear fragmentation and electromagnetic dissociation of $^{208}$Pb$^{82+}$ ions in collimators generates a flux of secondary fragments, which are lost in downstream dispersion suppressor and arc cells. These secondary ions may pose a performance limitation in upcoming runs since they can induce magnet quenches. In order to mitigate this risk, an alternative collimation technique, relying on bent crystals as primary collimators, will be used in forthcoming heavy-ion runs. In this paper, we study the power deposition in superconducting magnets by means of tracking and \textsc{FLUKA} shower simulations, comparing the standard collimation system against the crystal-based one. In order to quantify the predictive ability of the simulation model, we present absolute benchmarks against beam loss monitor measurements from the 2018 $^{208}$Pb$^{82+}$ run at 6.37~$Z$TeV. The benchmarks cover several hundred meters of beamline, from the primary collimators to the first arc cells. Based on these studies, we provide a detailed analysis of ion fragmentation and leakage to the cold magnets and quantify the expected quench margin in future $^{208}$Pb$^{82+}$ runs.
\end{abstract}

\maketitle

\section{Introduction}
\label{sec:introduction}

While pursuing a comprehensive proton physics program, the Large Hadron Collider (LHC) at CERN is also operated as a heavy-ion collider ($^{208}$Pb$^{82\textsc{+}}$)~\cite{JJowett2018}. The annual heavy-ion runs are  scheduled at the end of operational years and typically last for about one month. In the original LHC design, it was foreseen to collide 7~$Z$TeV $^{208}$Pb$^{82+}$ beams with a maximum stored energy of 3.8~MJ per beam. Already in the second physics run (2015-2018), this stored energy was significantly surpassed (13.3~MJ) because of a higher-than-nominal intensity (1.6$\times$10$^{11}$ $^{208}$Pb$^{82+}$ ions/beam)~\cite{Jowett2019,NFuster2020}. The beam energy achieved in Run~2 lead-lead runs was 6.37~$Z$TeV, i.e., almost the design value. The heavy-ion program will enter its High-Luminosity (HL) era in Run~3 (2022-2025), taking benefit from a further increase of the beam intensity to 2.2$\times$10$^{11}$ $^{208}$Pb$^{82+}$ ions/beam~\cite{RBruce2021}. The first run with these intensities is planned in 2023. A summary of the beam parameters is given in Table~\ref{tab:beamparam}. The ion energy in Run~3 will be 6.8~$Z$TeV, and might further increase to the nominal value of 7~$Z$TeV in Run~4 (2029-2032). Besides the $^{208}$Pb$^{82+}$ ions, lighter ion species (e.g. $^{16}$O$^{8+}$) are also considered for short runs in the future \cite{Dembinski:2020dam}. A beam test with $^{129}$Xe$^{54+}$ ions was conducted in 2017 \cite{Schaumann:2648699}.

\begin{table}[!b]
\caption{\label{tab:beamparam} Summary of $^{208}$Pb$^{82+}$ beam parameters in past LHC lead-lead runs~\cite{Jowett2019} and the present run~\cite{RBruce2021} (particle energy $E$, number of bunches $N_b$ per beam, bunch intensity $I_b$, stored beam energy $E_s=E N_b I_b$, bunch spacing $B_s$, and normalized transverse emittance $\epsilon_N$). For comparison, the first column shows the original design parameters \cite{JJowett2018}. For Run~1 and Run~2, we report the beam parameters, which gave the highest stored beam energy. The Run~3 parameters are the expected parameters for 2023-2025. Beyond Run~3, the beam intensity will be the same as in Run~3, whereas the beam energy could increase to 7~$Z$TeV.}
\begin{ruledtabular}
\begin{tabular}{lcccc}
  &Design&Run~1&Run~2&Run~3\\
 &&(2009-2013)&(2015-2018)&(2022-2025)\\
\hline
$E$ ($Z$TeV)& 7 & 3.2 & 6.37 & 6.8 \\
$N_b$& 592 & 338 & 733 & 1240 \\
$I_b$ ($10^{8}~\text{Pb}$)& 0.7 & 1.07 & 2.2 & 1.8\\
$E_s$ (MJ) & 3.8 & 1.9 & 13.3 & 19.9 \\
$B_s$ (ns) & 100 & 200 & 75 & 50 \\
$\epsilon_N$ ($\mu$m rad) & 1.5 & 2.0 & 2.3 & 1.65 \\
\end{tabular}
\end{ruledtabular}
\end{table}

In case of beam losses, even a small fraction of the stored beam energy can perturb the LHC performance by leading to magnet quenches, a phenomenon during which a superconducting (SC) magnets goes to normal-conducting state. Beam tests and theoretical models show that only 15-20~mW/cm$^3$ of energy deposited in the SC coils of a bending dipole is enough to induce a quench at 7~$Z$TeV~\cite{Auchmann2015,Bottura2019,Schaumann2020}. Therefore, LHC operation relies on multistage betatron and momentum collimation systems, which are indispensable for protecting the magnets against unavoidable beam halo losses~\cite{Coll_gene2}. The beam losses are continuously monitored by Beam Moss Monitors (BLMs), located all around the ring~\cite{BLMdes1,Dehning2007}. Continuous diffusion of beam particles from the beam core into the transverse tails is caused by a multitude of processes. Particle diffusion can for example be caused by the collisions in the interaction points or the interaction of the beams with electron cloud \cite{Petrov:1668162}. While these processes typically lead to slow but steady particle losses, other mechanisms can give rise to fast loss spikes, with rise times below one second. For example, sudden orbit oscillations were observed in the 2016 and 2018 heavy-ion runs ~\cite{Mirarchi2019,NFuster2020}, which resulted in recurring beam aborts by the BLMs. The orbit oscillations are believed to be induced by abrupt vibrations of certain magnets. The origin of these vibrations is still under study. 

\begin{figure}[!t]
\centering
\includegraphics[width=1\linewidth]{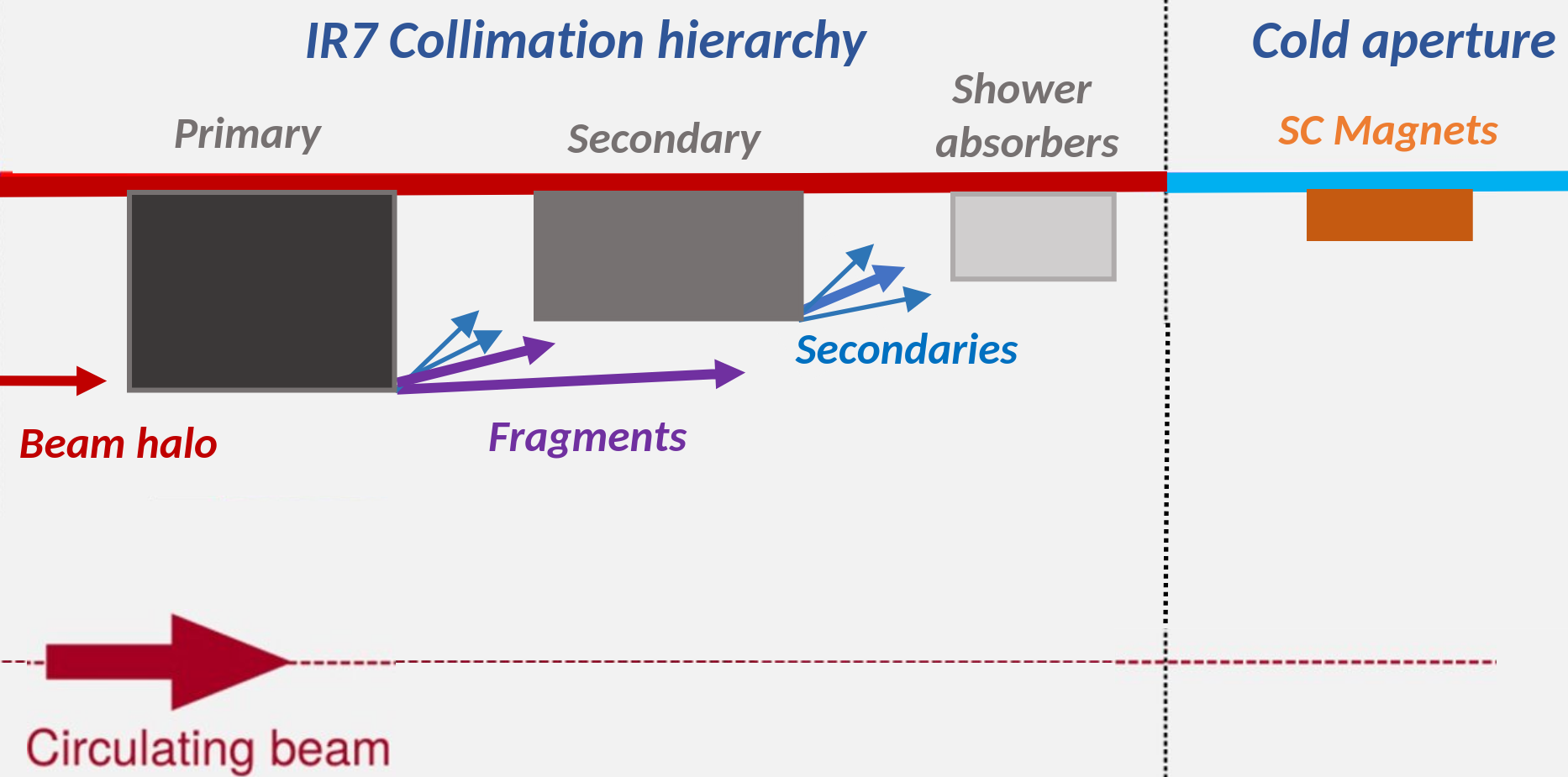}
\caption{Principle of the standard LHC collimation system. Figure inspired by Ref.~\cite{Aberle2020}.}
\label{Standard_collimation}
\end{figure}

The betatron and momentum collimation systems protect the machine against both slow particle diffusion and accidental beam losses. The collimation system of both beams are organized in a hierarchy of more than 100 collimators, which are placed at different transverse positions from the beam, as illustrated in Figure~\ref{Standard_collimation}. Most of the collimators are located in two insertion regions (IRs); IR7 hosts the betatron cleaning system and IR3 the off-momentum cleaning system. Operational experience showed that ion loss rates in the betatron system can reach higher peak values than in the momentum one. 

The betatron collimation system (IR7) exhibits a reduced cleaning efficiency in heavy-ion operation compared to proton runs due to the leakage of secondary fragments to downstream dispersion suppressor (DS) magnets~\cite{hermes16_nim}. These secondary ions are the result of hadronic fragmentation and electromagnetic dissociation in collimator blocks, mainly in the primary collimators, which are the first collimators intercepting beam halo particles in the collimation hierarchy. Due to the rising dispersion function in the DS, the fragments are lost in distinct lattice cells depending on their magnetic rigidity. 
The efficiency of the LHC collimation system proved to be sufficient in past heavy-ion runs, even when pushing the beam parameters beyond their design values. However, once the $^{208}$Pb$^{82+}$ ion energy and intensity increase further in upcoming runs, these fragments risk to induce magnet quenches in case the beam lifetime drops~\cite{Aberle2020}. 

A quench at top energy imposes a machine downtime of eight or more hours, during which the concerned SC magnets need to be brought back to their operational temperature (1.9~K for bending dipoles). Maintaining a good machine availability is an important aspect of LHC operation, in particular for the relatively short heavy-ion runs. Quench-induced downtimes would drastically reduce the availability for physics operation. Quenches can be prevented by setting sufficiently low beam abort thresholds on BLMs. However, frequent aborts by BLMs would also compromise the accelerator performance. As a design goal for the HL-LHC era, the betatron collimation system should allow for a beam lifetime of 0.2~hours over a period of ten seconds without risking a quench and without prematurely dumping the beams.

\begin{figure}[!b]
\centering
\includegraphics[angle=-90,width=0.8\linewidth]{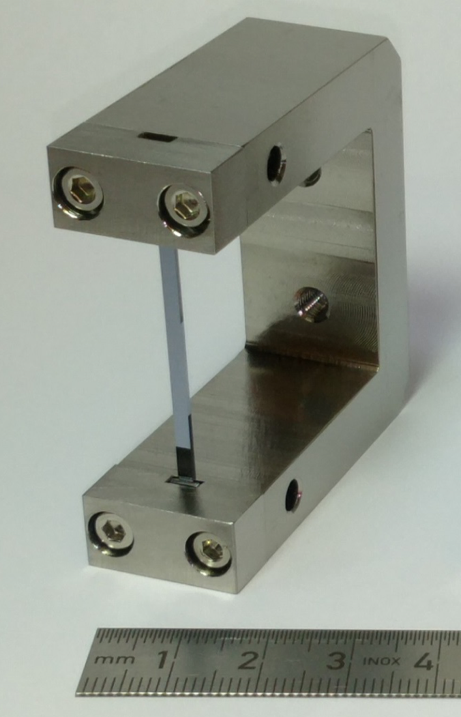}
\caption{Silicon strip crystal with its titanium holder.}
\label{Crystal_pic}
\end{figure}

As a primary solution to reduce the risk of halo-induced quenches in HL-LHC heavy-ion operation, it has been considered to substitute a dipole in the DS with shorter, but higher field magnets (11~T), creating space for an additional collimator~\cite{Aberle2020}. Presently, the installation of this assembly is, however, postponed.
As an alternative solution, a crystal-based collimation setup will be used featuring bent crystals of a few millimeters length~\cite{Aberle2020,DAndrea2021a}, as illustrated in Fig.~\ref{Crystal_pic}. Making use of the electromagnetic potential in their crystalline structures, bent crystals deflect halo particles through their atomic planes. This phenomenon, called channeling, can deviate incoming particles at angles of up to tens of microradians onto a downstream absorber. Due to the reduced probability of projectile fragmentation in the crystal and the large impact parameter on the channeled beam absorber, the crystal-based system reduces the fragment leakage to the DS and arc. So far, the crystal-based collimation setup has only been used during dedicated tests~\cite{Scandale2016,Rossi2017,Rossi2018,DAndrea2019,DAndrea2021,DAndrea2021a,Redaelli2021,Scandale2022} and in low-energy proton physics runs~\cite{Mirarchi2020}, but it will be employed in regular heavy-ion operation from 2023. The results of the run~II beam tests are summarized in Ref.~\cite{Dandrea_2023}.

Numerical simulations are indispensable for understanding and predicting the power deposition in coils of superconducting magnets. In order to assess performance limitations and quench margins, an advanced simulation chain has been developed at CERN for studying collimation losses~\cite{Skordis2018,Hermes_thesis}. The simulation chain couples the particle tracking code \textsc{SixTrack}~\cite{Schmidt1994,Robert2005,Sixtrck1}, updated to treat tracking of ions with different rigidities, with the Monte Carlo code \textsc{FLUKA}~\cite{Battistoni2015,Ahdida2022,FlukaWeb}. \textsc{FLUKA} is widely used for energy deposition studies in accelerator environments, in particular the LHC~\cite{Anton1}. 
Various BLM response studies were done in the past to validate the simulation setup for proton operation~\cite{bruce2014,Skordis2017,Anton1,ESkordis}. The physics processes are more involved for heavy ions, giving rise to a variety of ion fragments, which can leak to superconducting magnets. A first benchmark between simulated and measured BLM signals for heavy-ion beam collimation has been presented in Refs.~\cite{ESkordis,Skordis2017,NFuster2020}, based on data from the 2015 heavy-ion run. In this paper, we present a more comprehensive validation study, considering operational beam losses observed in 2018. Secondly, we present a first absolute BLM simulation benchmark for crystal-assisted collimation of $^{208}$Pb$^{82+}$ ion beams in the LHC. A model describing coherent effects of high-energy particles in crystals~\cite{Schoofs2014} has been recently incorporated into \textsc{FLUKA}~\cite{Ahdida2022}, making it possible to simulate a crystal-based setup through the same simulation chain. In this paper, we assess the ability of the crystal model to reproduce BLM measurements from controlled beam loss tests in the 2018 $^{208}$Pb$^{82+}$ run. Based on these benchmarks, we then use the simulation setup to study the secondary ion population leaking from the collimation system. Furthermore, we derive estimates of the power deposition in superconducting coils for HL-LHC beam parameters, comparing the standard collimation system with the crystal-assisted one. 

This paper is organized as follows. Section~\ref{sec:collimation} describes the LHC betatron collimation system, discusses the quench margin for betatron halo losses and presents the crystal collimation setup for future heavy-ion runs. Section~\ref{sec:simtools} describes the simulation framework. Sections~\ref{sec:benchmstdsystem} and \ref{sec:benchmcrysystem} present simulation benchmarks against 2018 measurements for the standard and crystal-based collimation systems, respectively. Based on those benchmarks, the fragment leakage to the cold aperture is detailed in Sec.~\ref{sec:frags}. Finally, Sec.~\ref{sec:powdep_coldmagnets} presents power deposition studies for the superconducting magnets for the HL-LHC era. A summary and concluding remarks are given in Sec.~\ref{sec:conclusions}.

\section{Betatron halo collimation}
\label{sec:collimation}

\subsection{The betatron collimation system}

\begin{figure*}[!t]
\centering
\includegraphics[width=0.98\linewidth]{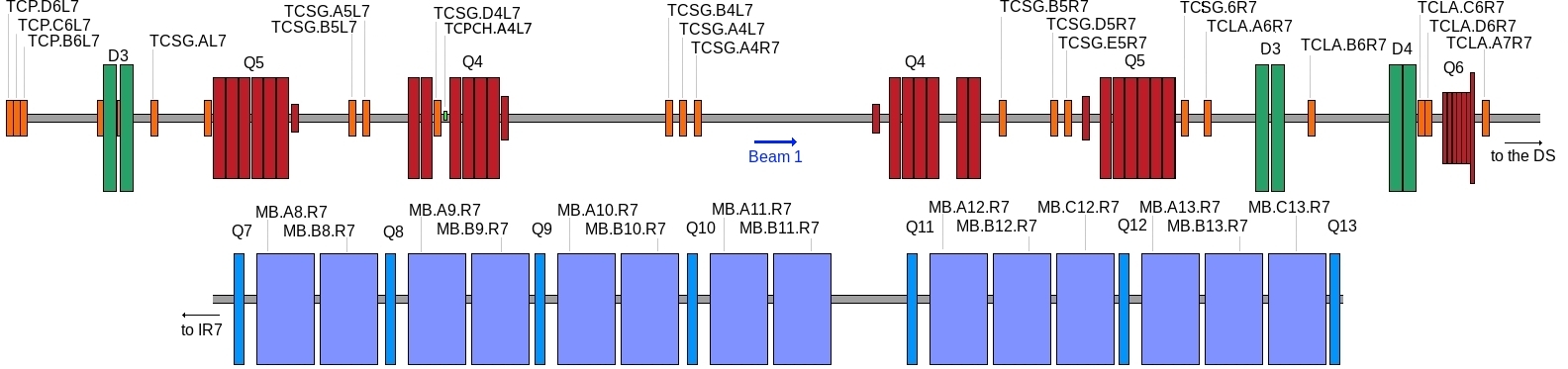}
\caption{Layout of the LHC collimation system for the clockwise-rotating beam, consisting of primary collimators (TCPs), secondary collimators (TCSGs) and shower absorbers (TCLAs) as well as the crystal on the horizontal plane (TCPCH). The downstream dispersion suppressor (cells 8-10) and the first arc cells (12-3) are shown. Dispersion suppressor cells are composed of two bending dipoles, one quadrupole and correctors (not shown).  Arc cells are longer, with one more dipole.}
\label{fig:colllayout}
\end{figure*}

The LHC multistage collimation system was designed to provide beam cleaning and passive protection against accidental beam losses \cite{Coll_gene2}. The betatron halo cleaning in IR7 is done with a three-stage collimator hierarchy; a similar hierarchy is adopted for off-momentum cleaning in IR3, but with fewer collimators. In this section, we describe only the betatron system since it is studied in this paper. For the standard collimation system, the primary collimators in IR7 (called TCP for Target Collimator Primary) are the closest elements to the beam in the entire ring; they are the first devices to intercept the beam halo at large betatron amplitudes. The tails of each beam are cleaned by three primary collimators of different azimuthal orientation (horizontal, vertical and skew). Secondary ion fragments and particle showers leaking from the primary collimators are intercepted by secondary collimators (called TCSG for Target Collimator Secondary Graphite) and by shower absorbers (called TCLA for Target Collimator Long Absorber), which are placed at a larger number of $\sigma$ from the beam. Like the primary collimators, TCSGs and TCLAs cover different planes. The layout of the collimation system in IR7 is illustrated in Fig.~\ref{fig:colllayout}. Additional collimators are installed in other regions, such as the tertiary collimators for local triplet-magnet protection in all experimental insertions (IR1, IR2, IR5 and IR8), collimators for physics-debris cleaning near the high-luminosity experiments in IR1 and IR5, and special ion collimators nearby IR2. All collimators have an active length of one meter, with the exception of the primary collimators in IR3 and IR7 and the ion collimators in IR2, which are 60~cm long.

The beam-intercepting components of collimators, called jaws, accommodate blocks of different absorber materials. Each collimator is composed of two opposite and movable jaws. In the first two LHC runs, the blocks of primary and secondary collimators were made of carbon-fiber composite (CFC) \cite{Coll_gene2}. In order to reduce the impedance budget of the collimators for proton operation in the HL-LHC era, some of the TCPs and TCSGs in IR7 were replaced before Run~3 by a new type of collimator with blocks made of molybdenum-carbide graphite (MoGR) \cite{Aberle2020}; MoGR has a reduced electrical resistivity compared to CFC. In addition, the new secondary collimators were coated with a thin molybdenum layer to further decrease the surface resistivity. The shower absorbers are made of a heavy tungsten alloy (Inermet-180)~\cite{Coll_gene2}. Similarly, the collimators in the experimental insertion regions have metallic absorber blocks (Inermet-180 or copper)~\cite{Coll_gene2}.

The LHC being a circular machine, beam halo particles can pass through a primary collimator multiple times before being subject to an inelastic nuclear collision or electromagnetic dissociation (the latter process being relevant for $^{208}$Pb$^{82+}$ ions, but not for protons). Some of the secondary particles can escape the collimation hierarchy and get lost outside the collimation system because of their different magnetic rigidity compared to beam particles~\cite{bruce2014,hermes16_nim}. In particular, secondaries can get lost on the cold aperture in the DS and arc immediately downstream of IR7 due to peaks in the dispersion function. The losses are clustered in odd-numbered half-cells, mainly in half-cells 9 and 11, and to a lesser extent in half-cell 13. The numbers reflect the position of the half-cells in the periodic lattice. The layout of the DS and first arc cells downstream of IR7 is illustrated in Fig.~\ref{fig:colllayout}. Dispersion suppressor half-cells are composed of two bending dipoles, one quadrupole and correctors. Arc half-cells are longer, with one dipole in addition. In proton operation, the particles leaking to cold magnets are mostly protons subject to single diffractive scattering in primary collimators. In heavy-ion operation, a variety of secondary ion fragments may leak to the cold region, resulting in a reduced collimation efficiency compared to proton runs~\cite{hermes16_nim}.

\subsection{Quench margin for betatron halo losses}

\begin{table*}[!t]
\caption{\label{tab:run2quenchtests} Overview of quench tests carried out with protons and $^{208}$Pb$^{82+}$ ions in Run~2. The table also lists the maximum power density in dipole coils ($(\Delta w/\Delta V)_{max}$) achieved in the tests. The power density values were reconstructed by means of shower simulations.}
\begin{ruledtabular}
\begin{tabular}{C{4cm}C{1.9cm}C{2.0cm}C{1.5cm}C{2.5cm}C{2.0cm}}
Loss term &Particle type (energy) & Time profile of loss rate & Quench & $(\Delta w/\Delta V)_{max}$ (reconstructed) & Complexity of simulations\\\hline
 &  & &  & & \\
Betatron collimation leakage from IR7 to DS~\cite{Salvachua2016} & protons (6.5~TeV) & Loss rate rising for 5~s& No & 20-25~mW/cm$^3$ \cite{ESkordis,leftthesis} & high\\
 &  & &  & & \\
Betatron collimation leakage from IR7 to DS~\cite{Hermes2016} & $^{208}$Pb$^{82+}$ (6.37~$Z$TeV) & Loss rate rising for 12~s & Yes &  25-30~mW/cm$^3$ \cite{ESkordis,leftthesis} & high\\
 &  & &  & & \\
Leakage of secondary ions from IP5 to DS~\cite{Schaumann2020}& $^{208}$Pb$^{82+}$ (6.37~$Z$TeV) & Constant loss rate for 20~s & Yes & 20~mW/cm$^3$ \cite{Schaumann2020} & low-medium \\
 &  & &  & & \\
\end{tabular}
\end{ruledtabular}
\end{table*}

A thorough understanding of magnet quench levels is essential for evaluating the risk of halo-induced magnet quenches in future heavy-ion operation. This concerns in particular the bending dipoles and main quadrupoles downstream of IR7. The quench level of a superconducting magnet is defined as the minimum power deposition density in the coils, which leads to a quench. Quench levels depend on intrinsic properties of SC magnets such as the cable and coil geometry and the type of superconductor~\cite{Auchmann2015}.
The bending dipoles and main quadrupoles in the LHC arcs and dispersion suppressors are built from Rutherford-type Nb-Ti/Cu cables. The coils are composed of inner and outer layers, each with radial width of 15~mm. A beam-induced quench is most likely to occur in the inner coils of a magnet, where the beam-induced energy deposition is the highest. The types of cables used in the inner coils of dipoles and quadrupoles differ in several parameters like filament size and number of strands. Other important variables for the quench level are the operating temperature (1.9~K for all arc and DS magnets), the local current density and magnetic field. Since the quench level decreases with increasing magnet current, the quench margin becomes tighter when operating the LHC at higher beam energies. In addition, quench levels vary as a function of the loss duration and time profile of beam losses~\cite{Auchmann2015}.
For slow losses, lasting several seconds or more, the quench level depends on the heat flow from the cables to the helium bath~\cite{Auchmann2015,Bottura2019}. In this case, the quench level is commonly expressed as the radially averaged power density in a cable since the heat can diffuse across the cable cross section during the loss period~\cite{Auchmann2015}. 

Dedicated beam tests with heavy-ion and proton beams were performed in past LHC runs to induce quenches in a controlled manner~\cite{Auchmann2015,Salvachua2016,Hermes2016,Schaumann2020}. The goal of these tests was to probe the quench level of magnets under realistic loss conditions. Table~\ref{tab:run2quenchtests} summarizes the quench tests carried out in Run~2. In two of the tests, beam losses were generated in the betatron collimation system in IR7 by means of the transverse feedback kicker; the first test was performed with 6.5~TeV proton beams \cite{Salvachua2016} and the second with 6.37~$Z$TeV $^{208}$Pb$^{82+}$ beams \cite{Hermes2016}. While no quench was observed in the proton beam test, a dipole quench was achieved with $^{208}$Pb$^{82+}$ ions. The quench occurred in cell 9 downstream of IR7, where the dispersion function has a first peak. The third test listed in Table~\ref{tab:run2quenchtests} was carried out in the DS next to the CMS insertion region (IR5), by steering secondary ions into a dipole using an orbit bump~\cite{Schaumann2020}. These secondary ions are the byproduct of collisions in the interaction point, from a process called bound-free pair production (BFPP)~\cite{Bruce2007BFPP}. In this process, one of the fully stripped $^{208}$Pb$^{82+}$ ions picks up an electron emerging from electron-positron pair production in an ultraperipheral collision. Since these ions are no longer fully stripped, they are lost in the downstream DS due to their different magnetic rigidity. By steering these secondary ions deep into one of the DS dipoles, a quench could be induced~\cite{Schaumann2020}. Although this test did not involve collimation losses, the test results still provide useful insight about the quench behavior of bending dipoles. 

Table~\ref{tab:run2quenchtests} also lists the estimated power deposition density in the dipole coils achieved in each of the three tests. The power deposition was reconstructed by means of tracking and particle shower simulations~\cite{ESkordis,leftthesis,Schaumann2020}, using the same simulation chain as in this paper. In the 2015 collimation quench test with $^{208}$Pb$^{82+}$ ions, the quench occurred at an estimated power deposition of 25-30~mW/cm$^3$. Prior to the quench, the $^{208}$Pb$^{82+}$ loss rate at collimators was rising for about twelve seconds. In case of a truly constant loss rate, the maximum acceptable power density without quenching is expected to be lower. This was indeed observed in the quench test with BFPP ions, where the quench occurred at an estimated power density of about 20~mW/cm$^3$~\cite{Schaumann2020}. In this case, the loss rate was constant for about 20~s before the quench developed. No quench was achieved in the collimation quench test with protons despite reaching a maximum power density of more than 20~mW/cm$^3$. This can possibly be explained by the different time profile of the loss rate. Like in the case of the ion collimation test, the loss rate was rising, but over a shorter period of only five seconds. 

Considering the complexity of the simulation setup for the collimation quench tests, some uncertainty remains concerning the exact quench level. In addition, the quench level will decrease further when increasing the beam energy in future runs. Based on these considerations, the quench level of dipoles is expected to be somewhat less than 20~mW/cm$^3$ at 7~TeV. Previous shower simulations showed that the power density in the coils of the most exposed dipoles can reach 40-60~mW/cm$^3$ if the beam lifetime drops to the design value of 0.2~hours in future higher-intensity heavy-ion runs \cite{Aberle2020}. Hence, magnet quenches can likely not be avoided in case of such lifetime dips without any upgrade of the collimation system.

\subsection{Crystal collimation}

\begin{figure}[!t]
\centering
\includegraphics[width=1\linewidth]{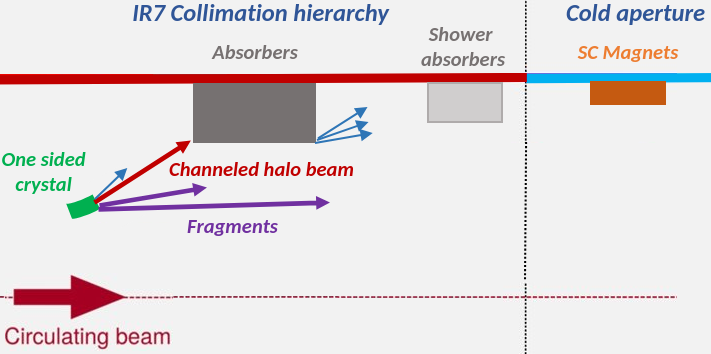}
\caption{ Working principle of the crystal collimation system. Figure inspired by Ref.~\cite{Aberle2020}.}
\label{Crystal_collimation}
\end{figure}

Crystal-assisted collimation is a novel collimation technique which will be used in forthcoming heavy-ion runs to reduce the leakage of secondary ion fragments to the cold magnets downstream of IR7. Crystal channeling is the coherent guiding or deflection of  positively charged particles by the lattice structure of a crystal; hence it is a phenomenon resulting from the extreme order in which the atoms are arranged in the crystalline lattice. Such crystals can be mechanically bent in order to impart a well defined curvature to the plane. Bent crystals can deflect heavy ions by tens of microradians, which is much more than the typical scattering angles achieved with amorphous materials.

Contrary to the standard multistage collimation system, which uses a chain of amorphous blocks placed at different gaps, crystal-assisted collimation relies on a bent crystal as a primary collimator. The channelled ions then impact on a secondary collimator in the betatron cleaning insertion and cannot make another turn in the machine. The principle of crystal-assisted collimation is illustrated in Fig.~\ref{Crystal_collimation}. Various bent crystals were previously tested in IR7, including silicon strip crystals and quasi-mosaic crystals. In particular, the efficiency of these crystals in reducing the leakage to the cold magnets has been assessed in dedicated tests with low-intensity proton and $^{208}$Pb$^{82+}$ ion beams~\cite{Scandale2016,Rossi2017,Rossi2018,DAndrea2019,DAndrea2021,DAndrea2021a,Redaelli2021,Scandale2022} before using crystal-assisted collimation in regular high-intensity operation in future heavy-ion runs. The installed crystals are single-sided devices intercepting the beam halo in either the horizontal or vertical plane, hence the tests were carried out for both cleaning planes, as well as for both counter-rotating beams. Figure~\ref{fig:colllayout} indicates the position of the horizontal  crystal in the IR7 layout (for the clockwise-rotating beam); in this case, the channeled beam impacts on the secondary collimator labelled TCSG.B4L7. In this paper, we focus mainly on this specific crystal setup.

\section{Simulation tools}
\label{sec:simtools}

We use a two-step modelling approach for simulating collimation losses and the resulting energy deposition in the machine. The simulation setup is based on the tracking code \textsc{SixTrack}~\cite{Schmidt1994,Robert2005,DeMaria2019} and the particle-matter interaction code \textsc{FLUKA}~\cite{Battistoni2015,Ahdida2022,FlukaWeb}. A coupling of the two codes has been developed previously at CERN~\cite{Skordis2018}, offering an advanced tool for studying multi-turn collimation losses. 
Tracking studies for crystal-assisted collimation in the LHC are discussed in detail in a separate paper~\cite{Cai2023}.
In the present paper, we investigate the resulting power deposition in machine equipment, which is computed in a second step using a stand-alone \textsc{FLUKA} model. The simulation chain is employed in this paper to compare the relative performance of the standard collimation system with the crystal-assisted one. In this section, we briefly summarize the relevant features and models of the simulation setup. We start with a short description of the relevant ion-matter interactions and the respective physics models.

\subsection{Ion fragmentation in collimators}

When ultra-relativistic $^{208}$Pb$^{82+}$ ions are intercepted by a collimator, they can fragment due to inelastic nuclear collisions or due to electromagnetic excitation in peripheral collisions~\cite{Flukabench3}. The latter process is known as electromagnetic dissociation (EMD). The electromagnetic forces can also lead to the break-up of target nuclei while the $^{208}$Pb$^{82+}$ projectile preserves its identity. While these collisions fall under the category of EMD processes, the resulting target fragments have a significantly different magnetic rigidity than the beam and will be intercepted by secondary collimators. As a result, these collisions not significant for the efficiency of the collimation system as the products are much less likely to leak to cold magnets.

With the standard collimation system in place, most of the $^{208}$Pb$^{82+}$ ions fragment in the primary collimators. Depending on the path length of ions in the material, the fragmentation process might only occur after multiple passages in the collimator blocks, i.e., after multiple turns in the machine. The likelihood, that the $^{208}$Pb$^{82+}$ ions are scattered to larger betatron amplitudes without fragmenting and are lost at other collimators in the hierarchy, is small. When using crystals as primary collimators, the probability of $^{208}$Pb$^{82+}$ fragmentation in the crystal itself is much reduced compared to the standard primary collimators since many ions are channeled and steered onto a secondary absorber. There are two aspects contributing to this reduced fragmentation probability in the crystal: first, and most important, the cumulative path length of ions in the amorphous regime is much shorter than in a standard bulk collimator as most of the ions are channelled; second, the collision cross section is reduced for ions subject to channeling. In both collimation schemes, most of the power deposition in cold magnets downstream of IR7 is due to ion fragments escaping from the primary beam-intercepting device, i.e., from the primary collimator or the crystal. Ions or other particles leaking from secondary collimators and shower absorbers yield a much smaller contribution (less than~2\%). 

\begin{table}[!t]
    \caption{Mean free path for inelastic nuclear and electromagnetic (EMD) collision processes of 7~$Z$TeV $^{208}$Pb$^{82+}$ ions in LHC collimator materials. The mean free path for EMD only considers projectile fragmentation. The last row gives the mean free path for $^{208}$Pb$^{82+}$ fragmentation independent of the process type, i.e., including both nuclear and electromagnetic fragmentation. The values originate from the \textsc{FLUKA} Monte Carlo code. The first row shows the material density. The assumed atomic fractions for MoGR are: 98.09\% carbon, 1.84\% molybdenum, 0.07\% titanium. }
    \label{tab:physical_processes}
    \begin{ruledtabular}
    \begin{tabular}{lccc}
          & {CFC} & {MoGR} & {Si}\\
         \hline
         $\rho$  & 1.67~g/cm$^3$ & 2.55~g/cm$^3$ & 2.33~g/cm$^3$ \\
         $\lambda_{inel}$ & 3.67~cm & 2.68~cm & 4.82~cm \\
         $\lambda_{EMD}$  & 26.34~cm & 10.57~cm & 7.50~cm\\
         $\lambda_{frag}$  & 3.22~cm  & 2.14~cm  & 2.94~cm  \\
    \end{tabular}
    \end{ruledtabular}
\end{table}

Table~\ref{tab:physical_processes} summarizes the mean free paths for nuclear fragmentation and electromagnetic dissociation of 7~$Z$TeV $^{208}$Pb$^{82+}$ ions in primary collimator materials (CFC and MoGR). The mean free path for EMD considers only projectile dissociation. The last row shows the effective mean free path for $^{208}$Pb$^{82+}$ fragmentation independent of the process type. Since the primary collimators are primarily composed of light nuclei (carbon), hadronic fragmentation of ions is much more likely than electromagnetic fragmentation. Compared to the CFC blocks used in Run~2, the relative importance of electromagnetic fragmentation increases for the primary collimator material in Run~3 (MoGR) due to the presence of molybdenum. The table also shows the corresponding mean free paths in silicon, which is the material used for crystal collimators. The table assume that the $^{208}$Pb$^{82+}$ ions experience the crystal as an amorphous absorber. The interaction cross sections are altered for projectiles subject to coherent effects in the crystal, as will be discussed later in this paper. In the amorphous regime, hadronic fragmentation dominates over electromagnetic fragmentation, although the relative importance of the latter significantly increases compared to the CFC and MoGR collimators.

A detailed description of the fragment production in collimators is essential for the simulation studies described in this paper. The fragment mass spectra are qualitatively different for hadronic and electromagnetic collisions, which can also affect the collimation efficiency. The relative difference of the mean free paths is hence not the only criterion for assessing the relative importance of the two processes for $^{208}$Pb$^{82+}$ collimation.
\textsc{FLUKA} integrates the \textsc{DPMJET-III}~\cite{DPMJET_proceeding,DPMJET_thesis} 
(Dual Parton Model and JETs) event generator for hadronic nucleus-nucleus collisions above 5~GeV/nucleon. EMD is simulated by means of the native nuclear collision generator in \textsc{FLUKA}, called \textsc{PEANUT}, which can handle photo-nuclear interactions~\cite{Flukabench3}. 
For both types of collisions, hadronic and EMD, the subsequent nuclear de-excitation, including evaporation, as well as fission, is treated by the same FLUKA module.
Benchmarks against experimental data showed that the models can accurately reproduce fragmentation yields following the impact of 400~$Z$GeV $^{208}$Pb beams on different fixed targets~\cite{Flukabench3}.

\subsection{Coherent effects in crystals}

Recently, a model describing coherent effects in crystals of positively charged hadrons with a momentum larger than 1~GeV/c ~\cite{Schoofs2014,Luigi_pap} has been incorporated into FLUKA~\cite{Ahdida2022}. The model enables the simulation of a crystal-based collimation setup and, as such, is a powerful tool to evaluate the expected performance of crystal-assisted $^{208}$Pb$^{82+}$ collimation in the LHC. 
 Particles entering the crystal at a transverse angle lower than the critical angle are channeled. The critical angle is defined as:
 \begin{equation}
     \theta_c=\sqrt{\frac{2U_0}{pv}}
     \label{eq:crit_angle}
 \end{equation}
where $U_0$ is the channel potential barrier, and $p$ and $v$ are the particle momentum and velocity, respectively. The trajectories of channeled particles are confined within adjacent potential barriers in the lattice until they either reach the end of the crystal or dechannel~\cite{Bible_crystal}. 
Dechanneling happens inside the crystal when a channeled particle undergoes a scattering event, which makes it escape the potential well. 
The suppression of close interactions experienced by channeled particles is implemented in FLUKA by means of a form factor~\cite{Schoofs2014}.

In case the projectile does not comply with the $\theta_c$ entry condition, it might be subject to volume reflection or volume capture. 
In the case of bent crystals, volume reflection applies to an unchanneled particle with an incident angle between $\theta_c$ and the crystal bending angle. The particle acquires a kick in the direction opposite to the crystal bending due to the plane's potential barrier. \textsc{FLUKA} will reflect a particle if its trajectory becomes tangent to the crystal planes. Instead of being reflected by the plane, such a particle can also undergo a scattering event that effectively captures it inside the channel: this corresponds to volume capture.

\begin{figure*}[!t]
\centering
\includegraphics[width=0.98\linewidth,trim={2cm 3cm 0 0cm},clip]{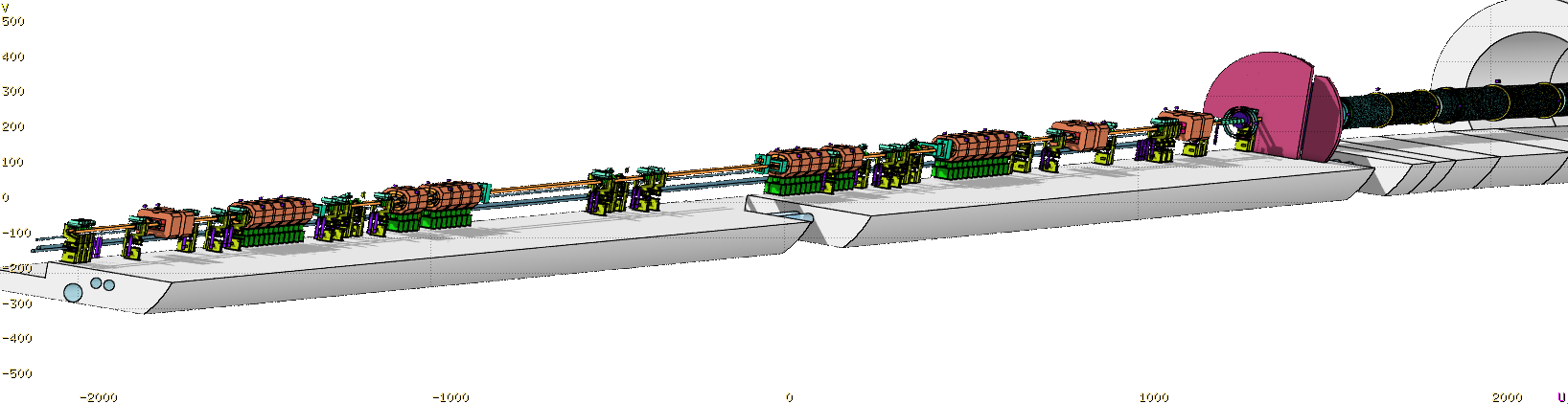}
\caption{Reproduction of the layout of the betatron collimation system for the clockwise-rotating beam (beam~1) in \textsc{FLUKA}. The figure shows the geometry from the primary collimators to the first half-cell of the downstream dispersion suppressor. The collimators from the counter-clockwise rotating beam (beam~2) are also included.}
\label{fig:colllayout_fluk}
\end{figure*}
 
\subsection{Simulation chain}

The simulation chain adopted in this paper consists of two independent steps. 
In the first step, the multi-turn loss distribution of ions on beam-intercepting devices in IR7 is calculated. This simulation step relies on the coupling between \textsc{SixTrack} for tracking ions in the magnetic fields of the machine and \textsc{FLUKA} for modelling ion interactions with collimators and the crystal. The initial impact parameter distribution of $^{208}$Pb$^{82+}$ ions on primary collimators or the crystal is not well known and might vary between loss events~\cite{NFuster2020}.
Details about the tracking setup can be found in Ref.~\cite{Cai2023}. We assume here as the initial condition that all $^{208}$Pb$^{82+}$ ions impact at a fixed distance from the primary collimator or crystal edge, i.e., at a fixed impact parameter $b$. The corresponding angle of the particles $\eta$ is determined by the matched beam optics. 

The impact parameter $b$ is typically assumed to be no larger than a few micrometers; the particle angles $\eta$ can reach a few tens of microradians~\cite{NFuster2020}. During one passage in the primary collimator, the effective path length in the block can be much shorter than the collimator length of 60~cm. This applies in particular to the case where the beam is focused in the cleaning plane and beam particles might only traverse the collimator tip. This concerns beam losses on the horizontal primary collimator, which are studied in this paper. Considering the mean free paths for nuclear or electromagnetic fragmentation, $^{208}$Pb$^{82+}$ ions can hence traverse the primary collimator unharmed in one passage. The same applies to $^{208}$Pb$^{82+}$ ions traversing the 4~mm short crystal. If an ion survives the passage and makes a full turn in the machine, it can impact at a different position on the same primary collimator or crystal. Secondary ion fragments emerging from a collimator or the crystal are only tracked with \textsc{SixTrack} if their energy is higher than 1~$Z$TeV. Other fragments are discarded~\cite{Cai2023}. If an ion touches the aperture of an element other than a collimator or a crystal, the tracking is also terminated. 
As particles pass through the surface of a collimator or a crystal during the tracking process, their positions and momentum are recorded for a second step. This includes both the surviving particles and the initial projectiles.

The second step of the simulation chain consists of using as source term the particles from the original pencil beam and its resulting multi-turn contribution for a \textsc{FLUKA} simulation of the full particle shower development. The simulation is carried out with a detailed \textsc{FLUKA} geometry model, spanning over several hundred meters of accelerator beamline. The model is described in more detail below. Contrary to the tracking simulations with the \textsc{SixTrack}-\text{FLUKA} coupling, low particle production and transport thresholds are used to obtain an accurate estimate of the energy deposition in machine elements and beam loss monitors. Following the fragmentation of a $^{208}$Pb$^{82+}$ ion in a collimator, \textsc{FLUKA} simulates the full cascade development, including the interaction of secondary fragments and the resulting hadronic and electromagnetic showers. The adopted production and transport thresholds  provide a good balance between accuracy and CPU time, as determined in previous energy deposition studies~\cite{ESkordis,Anton1}. Secondary ions were transported down to energies of 100~keV/n; similarly, hadrons and muons were tracked until their energies fell below 100~keV; the only exception were neutrons, which were transported down to thermal energies. The transport thresholds for electrons and positrons were chosen to be 1~MeV, and the thresholds for photons were 100~keV. These thresholds were applied in the entire geometry model, except for beam loss monitors, where lower cuts were needed (see Ref.~\cite{Anton1}). 

\subsection{Model for energy deposition simulations}

\begin{figure}[!t]
\centering
\scalebox{-1}[1]{\includegraphics[scale=0.24,trim={2cm 5cm 3cm 8cm},clip]{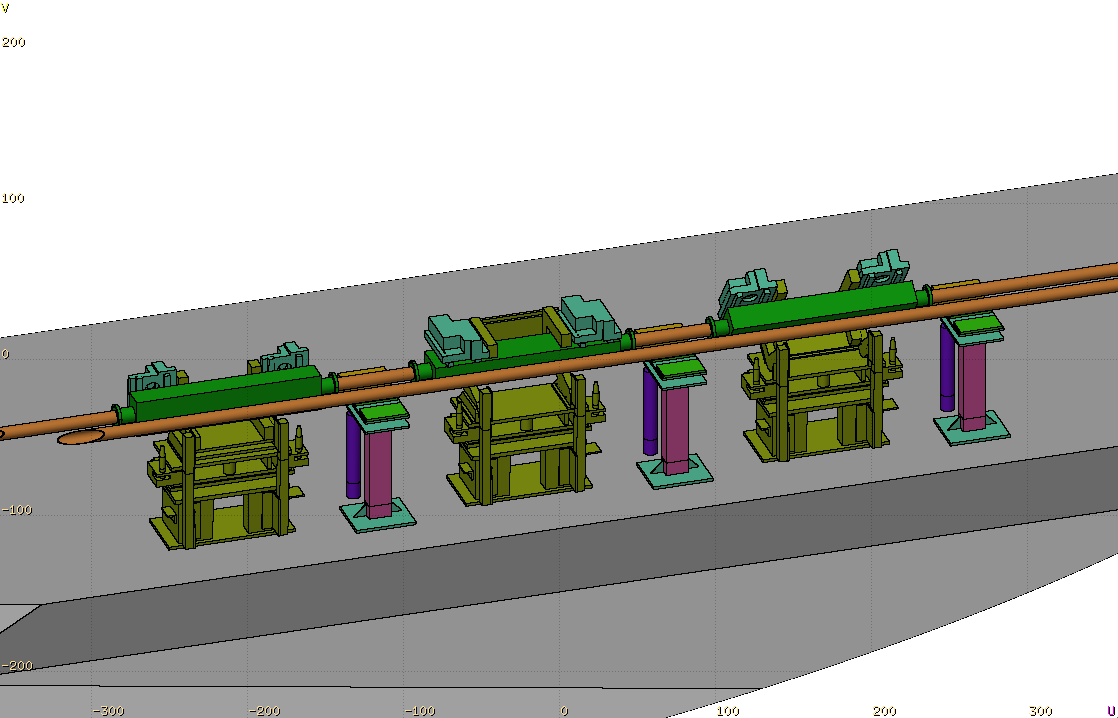}}
\caption{FLUKA geometry of IR7 collimators (in green), vacuum chambers (orange), beam support (pink) and BLM models (violet).}
\label{fig:FLUKA_LSS}
\end{figure}

In order to quantify the particle-induced energy deposition in the machine, a detailed geometry model of the accelerator is required, including key components such as beam-intercepting devices and magnets. Figures~\ref{fig:colllayout_fluk} and \ref{fig:FLUKA_LSS} illustrates the \textsc{FLUKA} geometry model of the insertion region and primary collimators used in this paper, respectively. Considering the sophisticated geometry of the LHC machine, a common repository of accelerator components was used to carry out the energy deposition studies. The FLUKA Element Database (FEDB)~\cite{FEDB} is a collection of \textsc{FLUKA} models of accelerator components such as magnets, collimators and beam loss monitors. The model of the accelerator line was assembled by means of the \textsc{LineBuilder} tool~\cite{FEDB}, which reproduces the lattice sequence as defined by \textsc{MAD-X} \cite{madxweb,Grote2002,Deniau2018}; it places different pieces of accelerator equipment from the FEDB into a master geometry model. The \textsc{LineBuilder} calculates the field strength of magnets according to the beam optics and beam energy. In addition, it automatically adjusts the gaps of collimators by taking into account the local value $\beta$-functions. Collimator gaps are usually expressed as multiples of the beam $\sigma$, which are provided to the \textsc{LineBuilder} as input parameters.  

A detailed account of the simulation uncertainty for LHC energy deposition simulations has been presented in Ref.~\cite{Anton1}. We recall a few of the relevant aspects here. The geometry models of collimators and magnets reproduce bulk structures, whereas some detailed geometrical features are approximated. This concerns, for example, the coils of superconducting magnets, which are modelled as a homogeneous material mixture of superconductor, stabilizer, insulator and liquid helium. This approximation is expected to allow for a reasonably accurate description of the particle shower development in the magnets since the radiation length is much longer than the dimensions of actual geometrical features. For instance, individual strands of superconducting cables have a diameter of about one millimeter, whereas the radiation length of Cu (stabilizer) or NbTi is more than one centimeter. Modelling the coils as a single material layer is hence considered justified for calculating the energy density distribution in the coils. 

The \textsc{FLUKA} geometry model also includes vacuum chambers and beam screens, which define the machine aperture. The aperture determines the particle loss distribution along the beam line, such as the loss positions of secondary fragments leaking from the collimators to the cold accelerator region. Because of manufacturing and alignment tolerances, the actual beam aperture can slightly differ from the ideal aperture assumed in the model. Hence, the actual loss positions can differ by up to several meters with respect to the simulation due to the grazing impact angles. In addition, the simulation model approximates the curvature of bending dipoles by means of one meter-long straight segments. This can locally distort the impact distribution on the aperture, but still gives a reasonably accurate estimate of the loss positions.

\section{Benchmark studies for standard heavy-ion collimation}
\label{sec:benchmstdsystem}

Particle showers induced by beam losses are continuously recorded by the LHC Beam Loss Monitor (BLM) system composed of almost 4000~ionization chambers~\cite{BLMdes1,Dehning2007}. The BLMs are installed near collimators, superconducting magnets and other accelerator equipment. The chambers are filled with N$_2$ gas and have a sensitive volume of about 1500 cm$^3$. Dose values are recorded with a 40~$\mu$s resolution, i.e., about twice per beam turn. Thresholds are set for each BLM individually to automatically trigger a beam dump in case of excessive losses. The BLM measurements are also essential for gaining a deeper understanding of beam loss mechanisms, for commissioning the collimation system, and for benchmarking simulations. BLMs are calibrated to provide absolute dose values, therefore allowing for a quantitative benchmark of simulation models.

The simulation chain described in the previous section has been benchmarked previously for standard  collimation cleaning (without crystals) by comparing simulated and measured BLM signals in IR7 and the downstream dispersion suppressor~\cite{Anton1,Skordis2017,leftthesis,NFuster2020}. Most of these studies were carried out for protons; the studies showed that the leakage of single diffractive protons to cold magnets in the dispersion suppressor is underestimated by a factor of three~\cite{Anton1,Skordis2017,leftthesis}. This discrepancy was mostly attributed to the absence of machine imperfections in the simulation model. The agreement could be improved by assuming an angular misalignment of the primary collimators~\cite{leftthesis}. The situation is more complex for $^{208}$Pb$^{82+}$ halo losses since a variety of secondary fragment species can leak to the cold region. A first quantitative comparison between simulated and measured BLM signals for $^{208}$Pb$^{82\textsc{+}}$ collimation losses was performed for the quench test at 6.37~$Z$TeV in 2015 (see also Sec.~\ref{sec:collimation}) ~\cite{Skordis2017,leftthesis,NFuster2020}. The simulation was found to underestimate BLM signals in the dispersion suppressor by up to a factor of five.

In this section, we compare new \textsc{FLUKA} shower simulations against BLM measurements recorded during the 2018 $^{208}$Pb$^{82+}$ run (6.37~$Z$TeV). In particular, we investigate fast beam loss events, which led to recurring beam aborts in 2018 operation. The characteristics of these loss events are described in the following subsection. The 2018 collimation settings were tighter than in 2015 and therefore the relative leakage to cold magnets was reduced compared to 2015. We assess the ability of the simulation model to reproduce the observed loss patterns for these settings, with particular attention to the dispersion suppressor. We also discuss the results dependence on the choice of initial conditions assumed in the simulations.

\subsection{Operational beam losses in 2018} 

\begin{figure}[!t]
\includegraphics[width=1\linewidth]{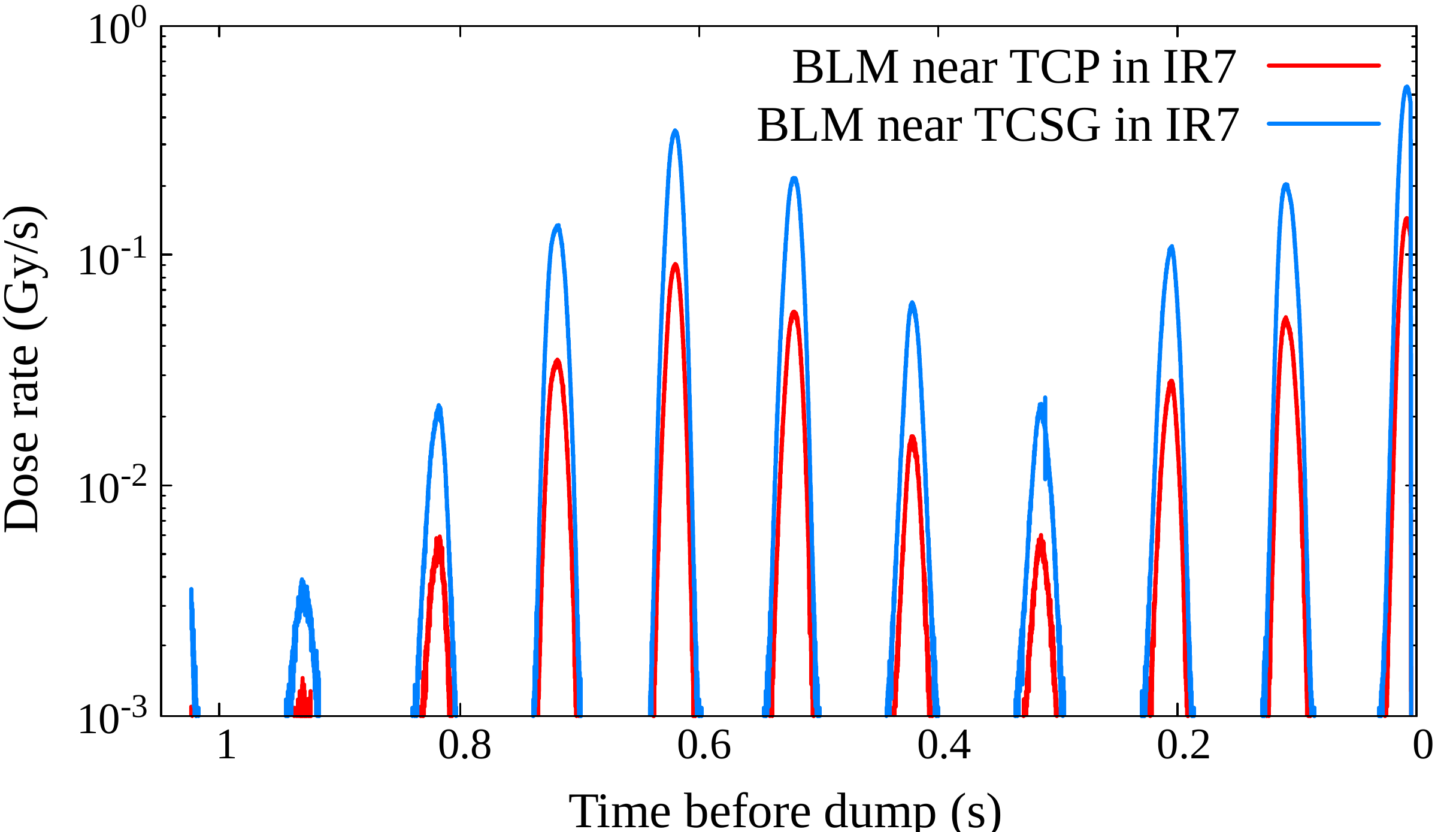}
\caption{Time profiles of a fast beam loss event measured with BLMs in Run~2 heavy-ion operation (28/11/2018). The BLMs were located in the IR7 collimation insertion, near a primary collimator (red curve) and a secondary collimator (blue curve).}
\label{10Hz_BLM}
\end{figure}

A series of fast beam loss events was observed during the 2018 heavy-ion run. All events occurred on the clockwise rotating beam (beam~1). The losses were believed to be caused by sudden vibrations of a magnet, which gave rise to orbit oscillations of the stored ion beam. The cause of the vibrations is still subject of investigation. The events led to peak loss rates in the betatron collimation system, which were up to a factor of 100~higher than in normal operation~\cite{Mirarchi2019}. The events consisted of multiple loss spikes, which repeated with a 8-12~Hz frequency and exhibited different amplitudes. Each spike lasted about 20-30~ms. Two BLM signals recorded during one of the events (28/11/2018) are shown in Fig.~\ref{10Hz_BLM}. 
In red is a BLM located near the primary collimator, which first intercepts the transverse beam tail. In blue is a BLM located further downstream at a secondary collimator, which intercepts secondary showers and fragments from upstream collimators. The same oscillations could be seen on all BLMs in the IR7 hierarchy of Beam~1. The events sometimes exceeded the BLM abort threshold after multiple oscillations, causing a protection dump during 6 out of the 48 physics fills~\cite{Mirarchi2019}. 
Similar events were also observed in the 2016 ion run as well as in 2018 proton operation. In these cases, the oscillations were induced at different locations.

\begin{figure}[!b]
\includegraphics[width=1\linewidth]{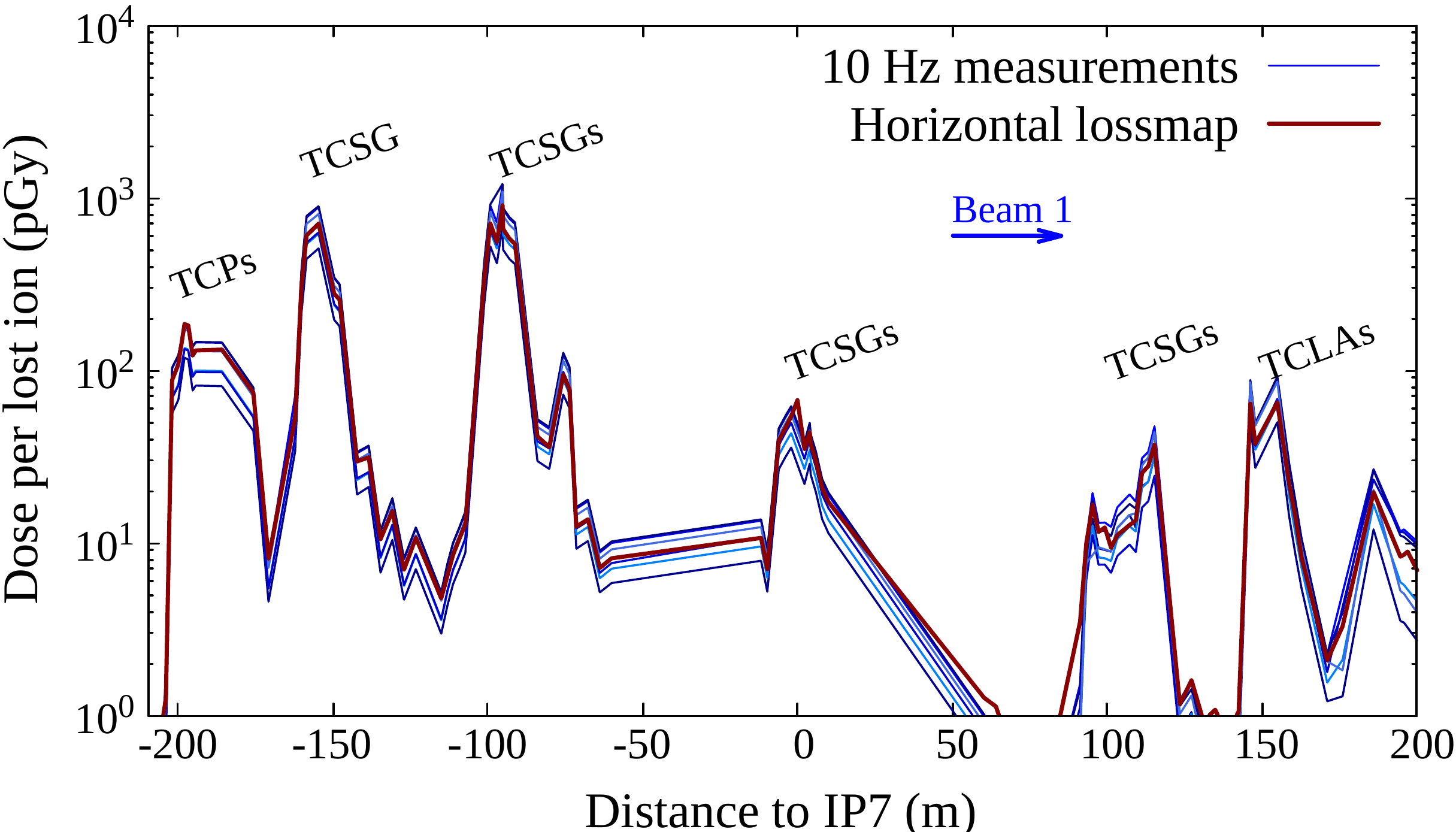} 
\caption{Measured BLM signal patterns along the IR7 betatron cleaning insertion (2018 $^{208}$Pb$^{82+}$ run). Fast beam loss events observed in operation (blue lines) are compared against a qualification loss map (red line). The latter was obtained by deliberately inducing betatron halo losses in the horizontal plane. The beam direction is from the left to right. The labels indicate the positions of primary collimators (TCPs), secondary collimators (TCSGs) and shower absorbers (TCLAs).}
\label{10Hz_B1H}
\end{figure}

Figure~\ref{10Hz_B1H} shows the spatial BLM dose patterns along the IR7 collimation system induced by the fast loss events in 2018 (blue curves). Each curve was normalized by the number of ions lost in the collimation system. The intensity loss was deduced from the beam current transformers (BCTs). The normalized curves exhibit some variation between events, which can likely be attributed to uncertainties in the normalization factors. The factors have an estimated error of a few tens of percent since the measured intensity loss was close to the achievable resolution of the BCTs. For comparison, the figure also shows a so-called qualification loss map (red curve). After shutdowns, the betatron cleaning system is qualified by intentionally creating losses in the horizontal or vertical planes using the transverse damper and low-intensity beams; the obtained BLM dose pattern along the collider ring, referred to as loss map, serves as a means of verifying the system hierarchy. The figure illustrates that the patterns of beam losses occurring during the fast loss events were qualitatively the same as the beam loss map for horizontal losses. This showed that the halo losses induced by the orbit oscillations were intercepted in the IR7 hierarchy as expected by design.

\subsection{Simulated and measured BLM signals}
\label{Chap:B1_Bench}

\begin{figure*}
\centering
\includegraphics[width=0.98\linewidth]{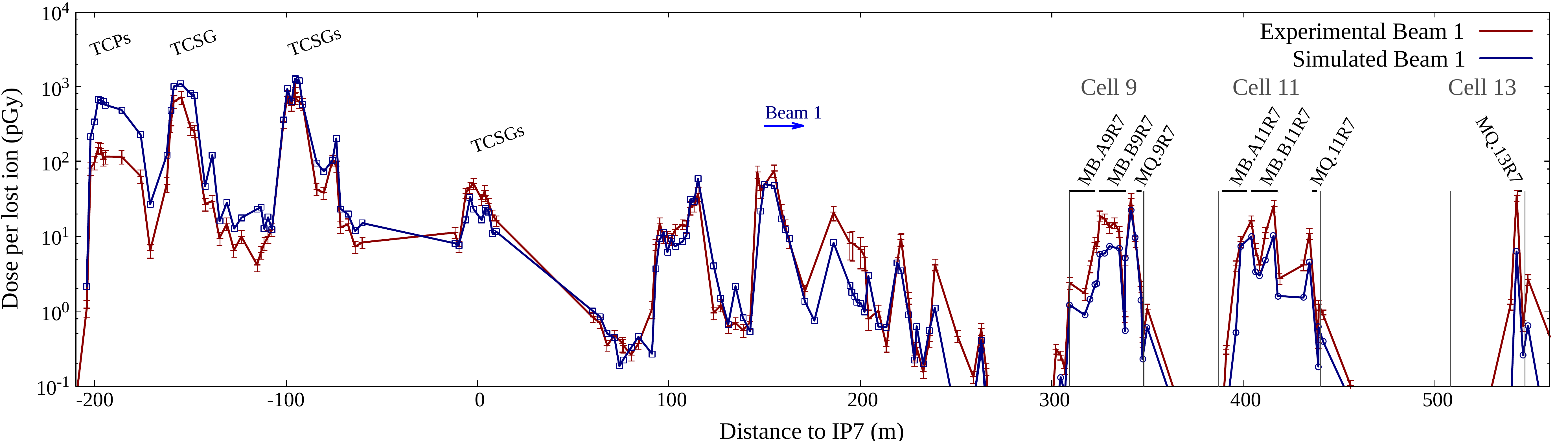} 
\caption{Simulated and experimental BLM signal patterns in the IR7 collimation insertion, the adjacent dispersion suppressor and the first arc cell. The signals are given per $^{208}$Pb$^{82+}$ ion intercepted in IR7. The measurements were averaged over the different loss events shown in Fig.~\ref{10Hz_BLM}. The beam direction is from left to right. The labels MQ, MB, TCP and TCSG represent quadrupoles, dipoles, primary and secondary collimators, respectively.}
\label{Beam1_benchmark}
\end{figure*}

In order to benchmark the simulation chain, BLM response simulations were performed for the 2018 machine configuration. In this year, asymmetric settings were used for the right and left jaw of the beam~1 horizontal primary collimator (5$\sigma$ and 5.5$\sigma$, respectively)~\cite{NFuster2020}. Here $\sigma$ is the transverse beam size for a normalized proton beam emittance of $\epsilon_{n}$=3.5~$\mu$m~rad. This asymmetric gap was necessary to reduce the leakage to a tertiary collimator. Because of the betatron motion, halo particles were first intercepted by the primary collimator jaw closer to the beam. Secondary collimators and active absorbers were symmetrically positioned at 6.5$\sigma$ and 10$\sigma$, respectively, as indicated in Table~\ref{tab:coll_settings}. For comparison, the table also lists the settings used in the 2015 $^{208}$Pb$^{82+}$ run. At 6.37~$Z$TeV, one $\sigma$ at the horizontal primary collimator corresponds to about 300~$\mu$m.

\begin{table}[!b]
    \centering
    \begin{tabular}{lccc}
    \hline\hline
         & 2015  operation & \multicolumn{2}{c}{2018 operation}\\
         &  Beam~1~\&~2 & Beam~1 &  Beam~2\\
         \hline
        TCP (H)  & 5.5 & 5.5(L)/5(R)   &  5  \\
        TCPs (V/S) & 5.5 & 5   & 5   \\
        TCSGs & 8 & 6.5 & 6.5 \\
        TCLAs & 14 & 10  & 10  \\
        \hline \hline
    \end{tabular}
    \caption{Collimator half gaps in IR7 during 2015 and 2018 heavy-ion operation (Run~2)~\cite{NFuster2020}. Values are expressed in units of beam $\sigma$, calculated for a normalized proton beam emittance of 3.5~$\mu$m~rad. The labels B1 and B2 specify the two counter-rotating beams. H/V/S refers to the azimuthal orientation of the primary collimators (horizontal/vertical/skew planes). L/R indicate the left and right jaw, respectively.}
    \label{tab:coll_settings}
\end{table}

In this study, we simulated only impacts on the horizontal primary collimator  since the beam oscillations in the described loss events were observed in the horizontal plane.
We use an impact parameter of 1~$\mu$m for $^{208}$Pb$^{82+}$ ions as initial condition, which yields the highest leakage of particles to the dispersion suppressor~\cite{NFuster2020}. 
The horizontal $\beta$-function decreases along the primary collimators and hence the direction of impacting halo particles points towards the beam center. Considering the small impact parameter, the $^{208}$Pb$^{82+}$ ions traverse only around five~centimeters of the absorber material since the jaws are not aligned with the beam envelope. If a $^{208}$Pb$^{82+}$ ion survives the passage through the absorber block and makes another turn in the machine, it can impact at a different position, i.e., at a different distance from the collimator edge. This spread of impact positions was accounted for by performing a two-step simulation as described in the previous section, i.e., multi-turn tracking simulations followed by shower simulations. 

Figure~\ref{Beam1_benchmark} compares the simulation results for IR7 and the adjacent cold region (blue curve) to measured BLM signals (red curve).

The measurements were averaged over the different loss events described in the previous section. The experimental error bars in the figure give the standard deviation of normalized signals from the different events.
The relative error of the measurements is about 5\%, which is considered satisfactory for this benchmark study. 
The statistical error of simulation results is a few percent for the highest signals in the IR, but can reach 20$\%$ in the dispersion suppressor due to the significant computational requirements.

An excellent agreement spanning a few orders of magnitude can be observed between simulated and experimental signals. Nevertheless, some discrepancies are found around the primary collimators (200~m upstream of IP7), in the DS (300-450~m downstream of IP7) and in the arc (from 450~m). As can be seen in the figure, the simulations overestimate the signals near the primary collimators by about a factor five. In order to assess the possible cause of this overestimation, we investigate the effect of different impact conditions as well as primary collimator tilts in the following subsection. In the DS (cells 9 and 11), the simulations are about a factor of two lower than the measurements. The factor of five underestimation at the Q13 magnet could be due to the very localized loss location making it very sensitive to any imperfections (e.g. aperture imperfections). All discrepancies found in this study were to some extent also observed in the previous simulation benchmark~\cite{ESkordis,NFuster2020} based on the controlled beam loss test in 2015. The underestimation in the DS was, however, larger in the previous study (factor of five compared to the factor of two observed in our study). This can partly be explained by the choice of a larger impact parameter in the previous simulation, which is known to reduce the losses in the DS~\cite{NFuster2020,Hermes_thesis}. Furthermore, collimator gaps were different, as shown in Table~\ref{tab:coll_settings}. In addition, the previous benchmark was performed for beam~2, which can be subject to a different leakage than beam~1.

\subsection{Dependence on initial conditions}
\label{Chap_Bench_imp}

\begin{figure}[h]
\includegraphics[width=1\linewidth]{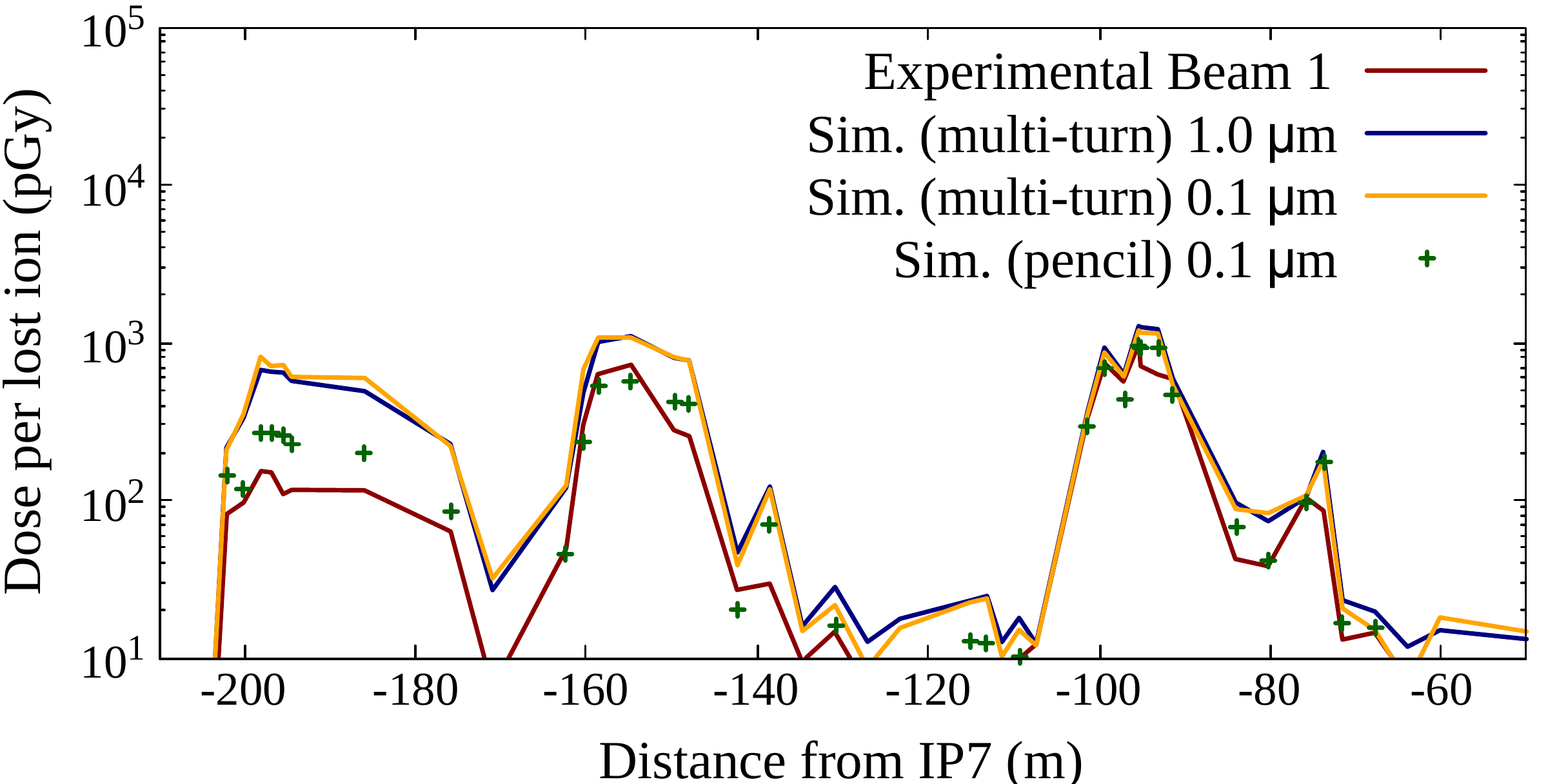}
\includegraphics[width=1\linewidth]{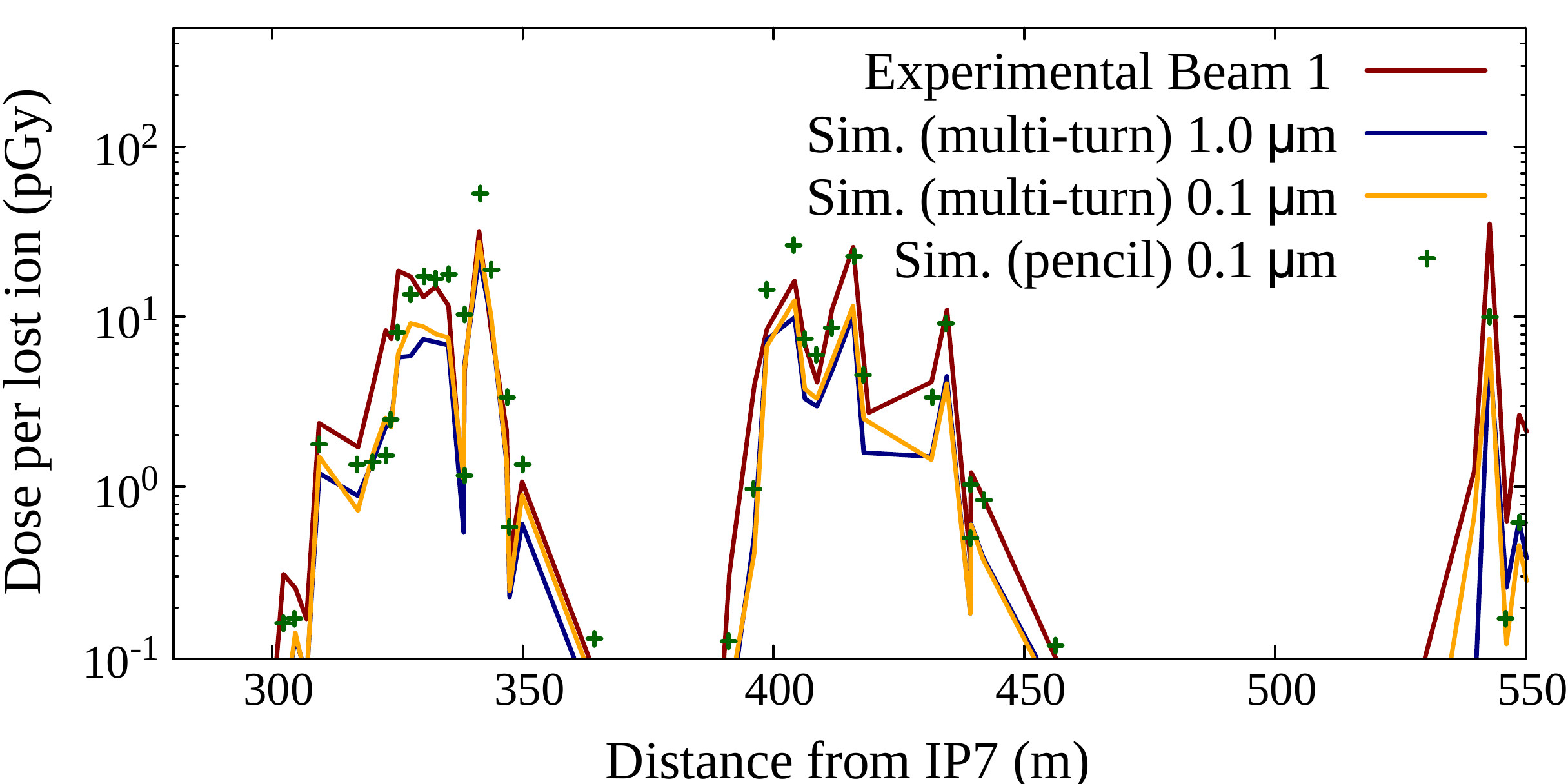}
\caption{Simulated and measured BLM patterns around the primary and the first secondary collimators (top) and in the dispersion suppressor and first arc cell (bottom). The yellow and blue curves take into account the spread of multi-turn impacts on the primary collimator, whereas the green crosses assume the non-physical case that all ions always impact on the same position.}
\label{Beam1_impacts}
\end{figure}

Several impact parameters in the sub-$\mu$m range were studied to investigate the sensitivity of BLM signals to beam loss conditions. Larger impact parameters were not considered here since they lead to reduced losses in the DS~\cite{NFuster2020}, and the simulation already underestimates those losses. Figure~\ref{Beam1_impacts} presents BLM patterns around the DS as well as the primary and first secondary collimators, considering impact parameters of 0.1~$\mu$m and 1~$\mu$m on the primary.

The results at the collimators show only a little dependence on the impact parameter below 1~$\mu$m.
The reason is that the impacts are smeared out by the multi-turn beam dynamics, considering that not all $^{208}$Pb$^{82+}$ ions will be subject to an inelastic nuclear collision or to electromagnetic dissociation during their first passage in the jaw. As a consequence, the ions can make one or more turns in the machine and can impact at a larger impact parameter with respect to the collimator edge in subsequent turns.
A lower initial impact parameter increases the number of turns after their first passage through the jaw and leads to a more diluted impact distribution. This increases the average impact parameter and therefore the BLM signals are very similar between the different cases.

The bottom graph in Figure~\ref{Beam1_impacts} shows the corresponding BLM signal patterns in the DS. Also in this case, the two different impact parameters yield similar results.
This result is compatible with previous tracking studies~\cite{NFuster2020}, where the leakage to cold magnets as a function of the impact parameter was studied. This study showed that the total amount of energy lost in cell 9 does not change significantly between 0.1 and 1~$\mu$m. By simulating the shower development in magnets, we find that BLM signals scale in a similar way. It is hence unlikely that the discrepancy between simulated and measured BLM signals can be explained by the choice of the initial impact parameter in the simulations.
It can, however, possibly explain why the measured BLM patterns of fast loss events are very similar to system qualification loss maps (see Fig.~\ref{10Hz_B1H}) even if the loss conditions might not have been exactly the same for the two cases.

Figure~\ref{Beam1_impacts} also shows a non-physical case (green dots), where the spatial spread of multi-turn impacts on the primary collimator is not simulated. Instead, $^{208}$Pb$^{82+}$ ions surviving the impact and making another turn in the machine were artificially loaded again at their initial position on the primary (0.1~$\mu$m from the edge). This study case shows smaller discrepancies of BLM signals near the primaries, the first secondaries and partially increases by a factor~2 the signals in the DS. This possibly indicates that the multi-turn spread of impacts on the right jaw of the primary collimator is overestimated in the other simulations (blue and yellow lines).

\section{Benchmark study for crystal-assisted heavy-ion collimation}
\label{sec:benchmcrysystem}

The first channeling experiments with 450~$Z$GeV and 6.5~$Z$TeV $^{208}$Pb$^{82+}$ beams in the LHC were performed in 2015 and 2016, respectively~\cite{Redaelli2021}. More extensive tests of the crystal-assisted collimation setup in IR7 were carried out in the 2018 heavy-ion run at 6.37~$Z$TeV ~\cite{DAndrea2019,DAndrea2021a}. Ions, which remain channeled throughout the entire crystal, receive a kick of several tens of microradians and impact on a certain secondary collimator that acts as the principal beam absorber. The objective of the tests in Run 2 was to assess  the system performance, before using crystal-assisted collimation in regular high-intensity operation in future heavy ion runs. The measured BLM patterns indicated a sizable reduction of the fragment leakage to superconducting magnets in the DS and arc, although a slight increase was observed around the first cold quadrupole (Q7) downstream of the collimation system~\cite{DAndrea2019,DAndrea2021a}. 

\begin{table}[!b]
    \centering
        \caption{\label{Coll_settings_crys}
        Collimator half gaps (Beam 1) in IR7 during 2018 heavy-ion operation (standard system)~\cite{NFuster2020} and during the beam test with the crystal-based system studied in this section~\cite{DAndrea2019,DAndrea2021a}. Values are expressed in units of beam $\sigma$, calculated for a normalized proton beam emittance of 3.5~$\mu$m~rad. H/V/S refers to the azimuthal orientation of the primary collimators (horizontal/vertical/skew planes). L/R indicate the left and right jaw, respectively. In the crystal test, different gaps were used for the secondary collimators upstream (TCSGs$_{upstream}$) and downstream (TCSGs$_{downstream}$) of the crystal (TCPCH).}
        \begin{ruledtabular}
    \begin{tabular}{lccc}
             & Standard system & Crystal system\\
                      & (2018 operation)  &  (2018 beam test)\\
        \hline
        TCP (H) & 5.5(L)/5(R) & 9 \\
        TCPs (V/S) & 5 & 9 \\
        TCSGs$_{upstream}$& 6.5 & 8.6 \\
        TCPCH & - & 5 \\
        TCSGs$_{downstream}$& 6.5 & 6.5\\
        TCLAs & 10 & 10 \\
    \end{tabular}
    \end{ruledtabular}
\end{table}

\begin{figure*}[!t]
\centering
\includegraphics[width=1\linewidth]{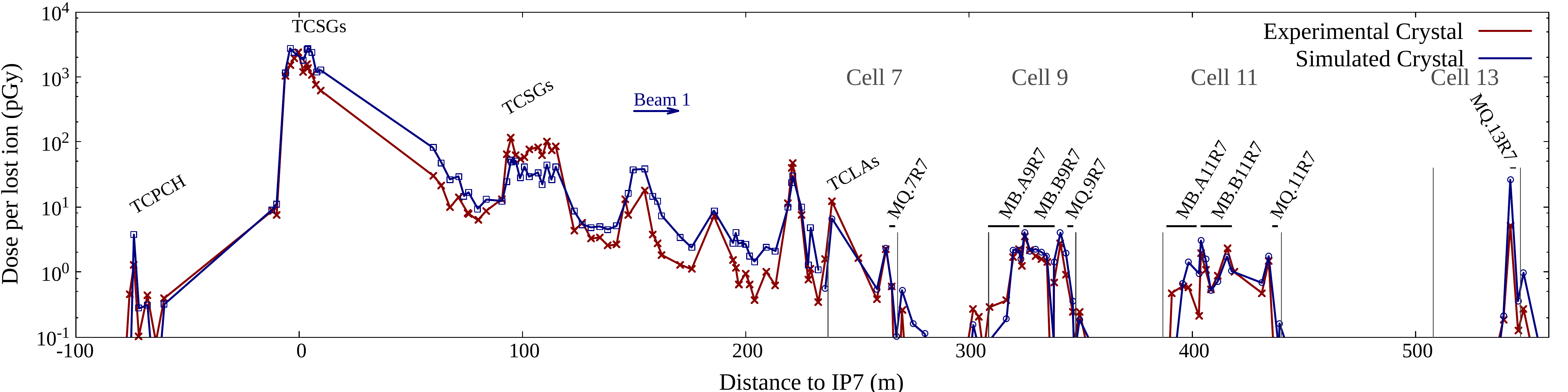} 
\caption{Simulated and experimental BLM signals for crystal-based $^{208}$Pb$^{82+}$ collimation in the betatron cleaning insertion. The experimental data are derived from a controlled beam loss experiment with 6.37~$Z$TeV beams in 2018. The beam direction is from left to right. The BLM signals are given per ion lost in the collimation system. The acronyms TCPCH, TCSG, TCLA, MB, MQ represent the crystal located on the horizontal plane, secondary collimators, shower absorbers, dipoles and quadrupoles, respectively.}
\label{Beam1_benchmark_cry}
\end{figure*}

In this section, we present a first absolute benchmark of the simulation chain for crystal-assisted ion collimation based on BLM data from 2018. The beam test considered for the benchmark was performed with the clockwise-rotating beam (Beam~1), using the collimator settings summarized in Table~\ref{Coll_settings_crys} (second column). For completeness, the table also shows the operational settings used for the standard system in 2018. The test was carried out by inserting a 4~mm long silicon strip-crystal in the horizontal plane. The crystal with a 65~$\mu$rad bending angle was installed on the external side of the machine, about 130~m downstream of the regular primary collimators. The primary collimators, as well as secondary collimators upstream of the crystal, were retracted compared to their nominal position. The secondary collimators downstream of the crystal  and the active shower absorbers were maintained at their nominal gap used in 2018 operation. The secondary collimator intercepting the channeled ions (named TCSG.B4L7 in Fig.~\ref{fig:colllayout}) was located about 70~m downstream of the crystal position, approximately in the center of the insertion region. The position of the crystals was carefully chosen such that the phase advance to existing absorbers was about $\pi/2$~\cite{Mirarchi2017}.

In analogy to the simulation benchmark for the standard collimation system, it was assumed that the $^{208}$Pb$^{82+}$ halo particles impact at certain distance from the crystal edge. The initial impact parameter was chosen to be 1~$\mu$m. Details about the multi-turn tracking studies for this benchmark can be found in Refs.~\cite{Cai2022,Cai2023}. Coherent effects in the crystal were simulated with the model described in Sec.~\ref{sec:simtools}. The probability of inelastic nuclear collisions is greatly reduced for ions subject to channeling. In the simulation model, this is accounted for by considering the average nuclear density experienced by a particle oscillating between neighboring crystal planes~\cite{Ahdida2022}. The model was found to reproduce well the experimentally observed reduction of nuclear collisions for channeled protons~\cite{Ahdida2022}. Theoretical considerations indicate that for ions the probability of EMD is also reduced~\cite{Scandale2013}. For 7~$Z$TeV $^{208}$Pb$^{82+}$ beams, Ref.~\cite{Scandale2013} suggests that the probability for projectile dissociation is as low as 3$\times$10$^{-5}$/mm during channeling. For amorphous silicon, this probability is significantly higher (about 10$^{-2}$/mm). 

No experimental data exists, which quantifies the reduction of projectile dissociation of channeled ions in this energy regime. In order to assess the impact of EMD on the simulation benchmark, we studied the two extreme cases, where EMD is either fully suppressed in channeling or retains the same probability as in the amorphous regime. The results indicate that dissociation of channeled ions yields only a small contribution to the fragment population lost in cold magnets. Since channeled ions receive an angular kick, the dissociation products are less likely to escape the collimation system. The benchmark results for the cold regions are therefore largely independent of the model assumption concerning EMD in channeling. We will present a detailed assessment of fragment production and leakage in the next section, where this aspect will also be discussed. Hereafter, we assume that projectile dissociation is not suppressed in channeling.

Figure~\ref{Beam1_benchmark_cry} presents an absolute comparison between BLM simulations and measurements for the considered beam test at 6.37~$Z$TeV. The BLM dose values recorded during the test (red curve) were normalized by the number of lost ions using fast beam current transformer measurements. The simulation results are shown by the blue curve. The statistical error of simulated BLM signals is at most a few percent for the highest signals in the insertion region, but can be up to 20$\%$ in the DS and the arc. The overall agreement between simulation and experiment is found to be remarkably good. The simulation reproduces well the measurement pattern over several orders of magnitude for more than 130 monitors distributed over 700~m of beam line. In particular, the BLM signals around the secondary collimator, which intercepts channeled beam (at s~$\textsc{=}$~-7~m in Fig.~\ref{Beam1_benchmark_cry}), match well. This indicates that the fraction of channeled ions can be well predicted by the model. A factor~3 discrepancy is observed at some downstream collimators, with some of the BLM signals being underestimated and some overestimated by the simulation. On the other hand, the BLM pattern provides an even better agreement in the DS compared to what was achieved for the simulation of standard heavy-ion collimation in the previous section. The simulated BLM signals and loss patterns in cell 9 and 11 are very close to the measurements, giving confidence that the \textsc{FLUKA} model accurately reproduces the power deposition inside the superconducting magnets in both cells. The agreement is less good for cell~13, where the simulation overestimates measured signals by a factor of five. As will be discussed in the next section, the losses in this cell are dominated by one single isotope ($^{206}$Pb$^{82+}$).

\section{Fragment leakage to cold magnets}
\label{sec:frags}

A first simulation study of ion fragments leaking from the standard collimation system to cold magnets has been presented in Refs.~\cite{hermes16_nim}. Although the reinteraction of fragments in collimators was neglected in this study, it could show that a variety of different secondary ion species contribute to the energy deposition in the loss clusters in the DS and arc. In this section, we provide a detailed assessment of the fragment production and leakage for the crystal-based system. The differences with respect to the standard system are discussed. The presented results correspond to same machine configuration as in the two previous sections, i.e., 2018 collimator settings and a beam energy of 6.37~$Z$TeV. In addition, we present some considerations and results for future runs. In all cases, we consider beam halo losses in the horizontal plane (clockwise-rotating beam) and we assume that the ions have an impact parameter of 1~$\mu$m in the first turn, based on~\cite{NFuster2020}. 
Although the presented results correspond to a specific crystal configuration, similar conclusions are expected for the vertical plane and the counter-rotating beam because the dispersive nature of losses in the DS is qualitatively similar independently of where fragments are produced. 

\subsection{Secondary fragment production}

\begin{table}[!b]
\caption{\label{tab:fragprim}Simulated fraction of $^{208}$Pb$^{82+}$ ions (6.37~$Z$TeV), which are subject to electromagnetic or hadronic fragmentation in the primary collimator (standard system) or the Si crystal (crystal-assisted system). The table also shows if the fragmentation took place while the $^{208}$Pb$^{82+}$ was channeled or not channeled in the crystal. The results correspond to 2018 operational settings, assuming losses in the horizontal plane (clockwise-rotating beam).
}

\begin{ruledtabular}
\begin{tabular}{lccc}
     & Regular system &\multicolumn{2}{c}{Crystal-based system}\\
     &  & (channeled) & (not chann.)\\
     \hline
EMD      & 0.120 & 0.038      & 0.042      \\
Hadronic & 0.876 & 0.006      & 0.091      \\
\end{tabular}
\end{ruledtabular}
\end{table}

Table~\ref{tab:fragprim} summarizes the fraction of $^{208}$Pb$^{82+}$ ions subject to electromagnetic and hadronic fragmentation in the primary collimator (standard system) and the crystal (crystal-based system). The table does not consider EMD of target nuclei, where the $^{208}$Pb$^{82+}$ ions preserve their identity. For the case of the standard system, a large fraction of $^{208}$Pb$^{82+}$ ions (about 88\%) break up in inelastic nuclear collisions in the primary collimator. Since the collimator is made of a light material (carbon), EMD is less likely (see also Table~\ref{tab:physical_processes}). In total, more than 99\% of the lost $^{208}$Pb$^{82+}$ ions are subject to fragmentation in the primary collimator, whereas the rest breaks up in other collimators. For the crystal-based system, the simulation predicts that less than 18\% of the $^{208}$Pb$^{82+}$ ions undergo fragmentation in the crystal, while the rest impacts on the secondary collimator, which acts as channeled beam-absorber. The table also shows whether ions were in the channeling regime when the fragmentation occurred in the crystal.
Like in the previous section, it was assumed in the \textsc{FLUKA} simulation that the EMD cross section is not reduced for channeled ions. In this case, the fraction of electromagnetic fragmentation in the channeling regime is comparable to the non-channeling regime, affecting about 4\% of the ions in both cases. The slight difference is due to the different cumulative path length of the projectile  in the different regimes. The fraction of $^{208}$Pb$^{82+}$ ions subject to nuclear fragmentation in the non-channeling regime is about 9\%, i.e., about double as high as electromagnetic fragmentation. The relative importance of the two different processes is more balanced than in the carbon collimators since the difference of hadronic and EMD cross sections becomes smaller in the silicon crystal. 

\begin{figure}[!t]
\centering
\includegraphics[width=\linewidth,trim={0 0 0 0},clip,]{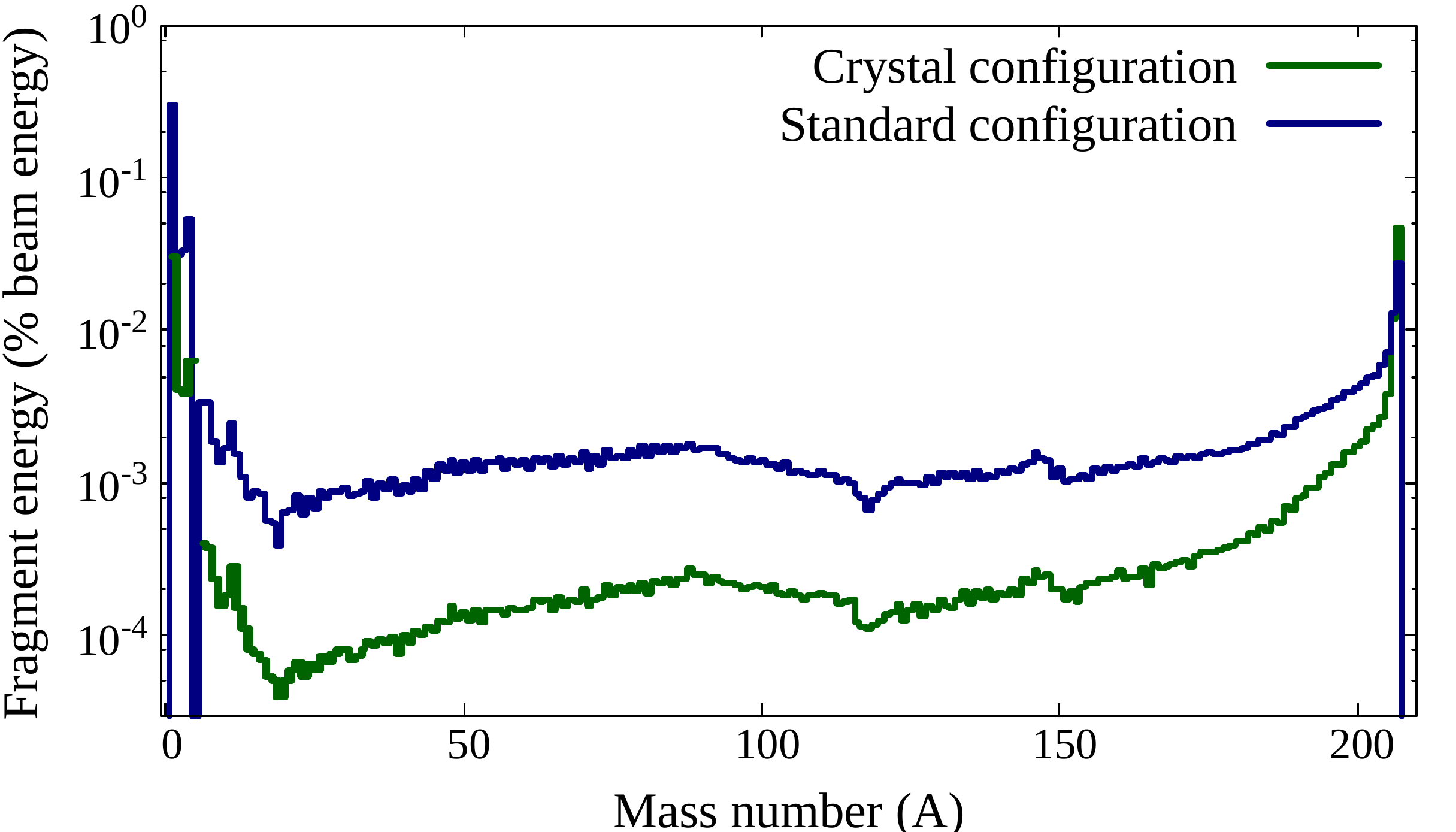}
\includegraphics[width=\linewidth,trim={0 0 0 0},clip,]{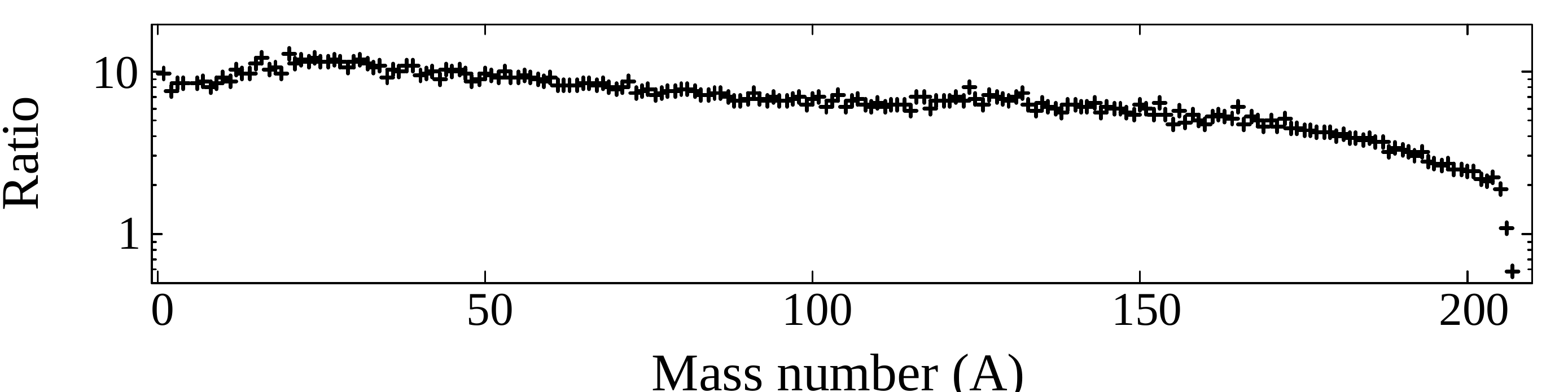}
\caption{Energy fraction carried by secondary ion fragments escaping from the primary collimator or the crystal (top figure). The results are given as a function of the mass number. Only fragments with kinetic energies above 1~TeV per nucleon were considered. The results correspond to 2018 machine settings (6.37~$Z$TeV). The ratio of energy fraction carried by the secondary ion fragments between the two collimation techniques is illustrated in the bottom figure.}
\label{fig:energy_leaving}
\end{figure}

The secondary fragments produced by $^{208}$Pb$^{82+}$ ion interactions with the collimator material can escape from the primary collimator jaws, or they might be subject to further interactions in the  collimator blocks depending on their path length in the material. Fragments created in the silicon crystal have a smaller chance to re-interact due to the much smaller crystal length. The ion species leaking from the primary collimator or crystal comprise a large range of mass numbers, from hydrogen up to lead. Figure~\ref{fig:energy_leaving} illustrates the relative energy fraction carried by different secondary fragments escaping from the primary collimator and the crystal, respectively. The energy is expressed as a fraction of the kinetic energy of primary $^{208}$Pb$^{82+}$ ions (6.37~$Z$TeV). Only high-energy ion fragments ($>$1~TeV/n) were considered. Lower-energy fragments, as well as other secondary particles are mainly lost before reaching the cold apertures.
Although the number of $^{208}$Pb$^{82+}$ ions fragmenting in the crystal is much less than in a regular primary collimator, the figure shows a higher abundance of fragments with $A$=207 escaping from the crystal. This can be attributed to the reduced re-interaction rate of these fragments in the crystal compared to the standard collimators. For other species, the energy fraction carried by fragments is up to one order of magnitude less in the crystal-based system than in the standard system.

\begin{figure}[!t]
\centering
\includegraphics[width=\linewidth,trim={0 0 0cm 0},clip,]{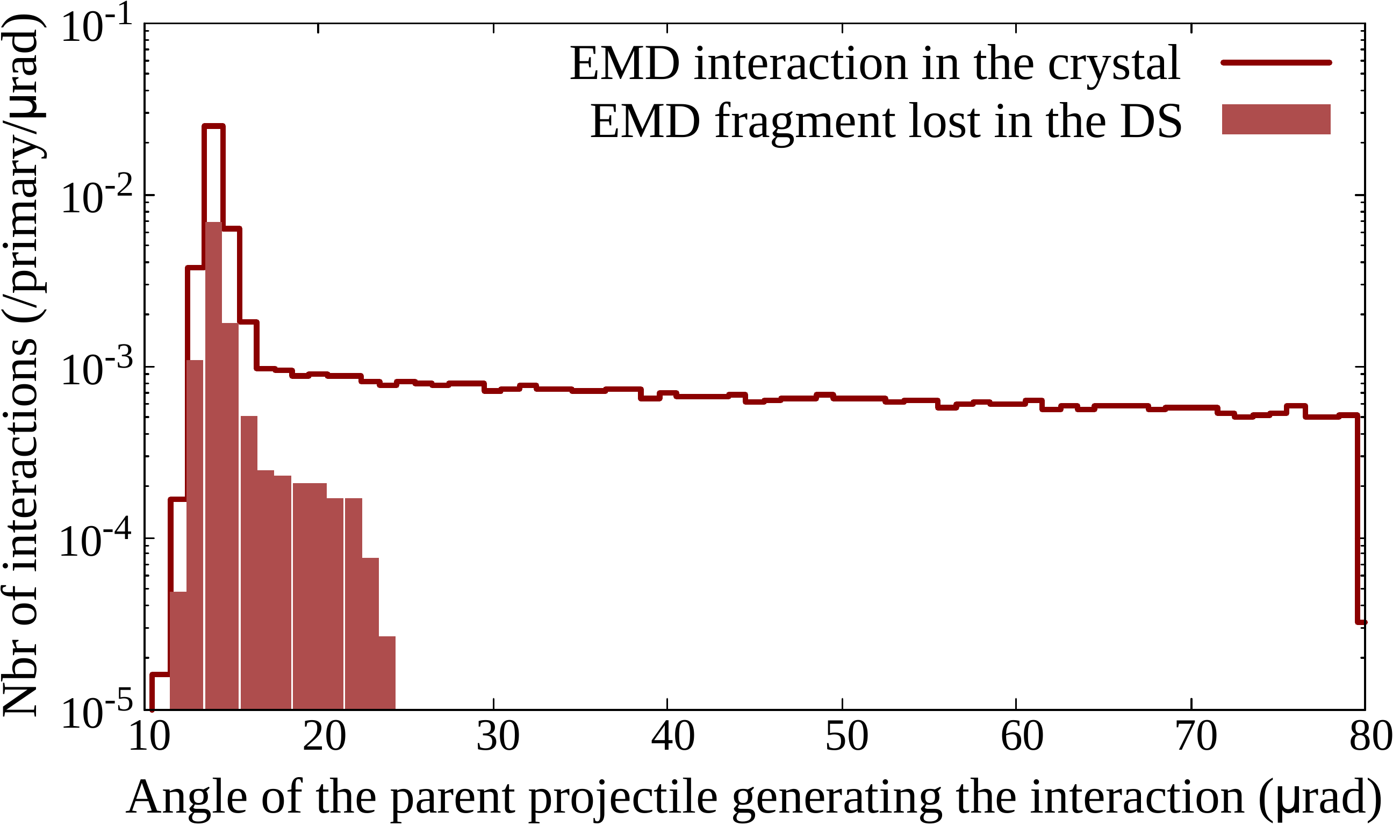}
\caption{Simulated angular distribution (horizontal plane) of $^{208}$Pb$^{82+}$ ions at the moment of electromagnetic fragmentation in the crystal. The line includes all $^{208}$Pb$^{82+}$ electromagnetic fragmentation interactions in the crystal, whereas the filled histograms include only the subset of collisions resulting in the leakage of high-energy secondary fragments ($>$1~TeV/n) to the cold magnets downstream of IR7. All values are expressed as a fraction of the total number of $^{208}$Pb$^{82+}$ halo ions intercepted in the betatron collimation system. The results correspond to 2018 machine settings (6.37~$Z$TeV).}
\label{fig:interaction_angle}
\end{figure}

The simulations show that a large fraction ($>$98\%) of the power deposition in cold magnets downstream of IR7 is due to secondary ions emerging from the primary collimator (standard system) or from the crystal (crystal-based system). The contribution of secondary ions from other collimators is small. In order that a fragment from the primary or crystal can leak to the dispersion suppressor or arc, its magnetic rigidity must be close enough to the beam rigidity, otherwise it would be lost in IR7~\cite{hermes16_nim}. As a consequence, only selected fragments in Fig.~\ref{fig:energy_leaving} contribute to the power deposition in cold magnets. Another factor determining whether a particle can reach the cold magnets is the particle direction. This concerns in particular fragments emerging from the crystal. The simulation shows that the leakage of fragments is much reduced if the primary $^{208}$Pb$^{82+}$ ion was channeled and therefore received a horizontal kick prior to the collision. 

This is illustrated in Fig.~\ref{fig:interaction_angle}, which presents the angular distribution of $^{208}$Pb$^{82+}$ ions at the moment of electromagnetic fragmentation in the crystal, as well as the subset of interactions where fragments leak to the cold magnets. The peak around 14~$\mu$rad corresponds mainly to ions not subject to channeling (14~$\mu$rad is the angle $\eta$ of the incoming projectiles). These ions suffer only a small angular deviation from their initial direction due to multiple scattering. The long tail up to an angle of 80~$\mu$rad corresponds to channeled ions. The larger the kick received by the $^{208}$Pb$^{82+}$ ions, the less likely becomes the leakage of secondary fragments to the cold region. The contribution drops steeply for projectile angles larger than 24~$\mu$rad.
The results therefore suggest that the efficiency of the crystal-based system is rather independent of the actual suppression of electromagnetic collisions in channeling. Similar conclusions can be drawn for hadronic collisions.

\subsection{Fragment loss distribution on the cold aperture}

\begin{figure}[!t]
\includegraphics[width=1\linewidth]{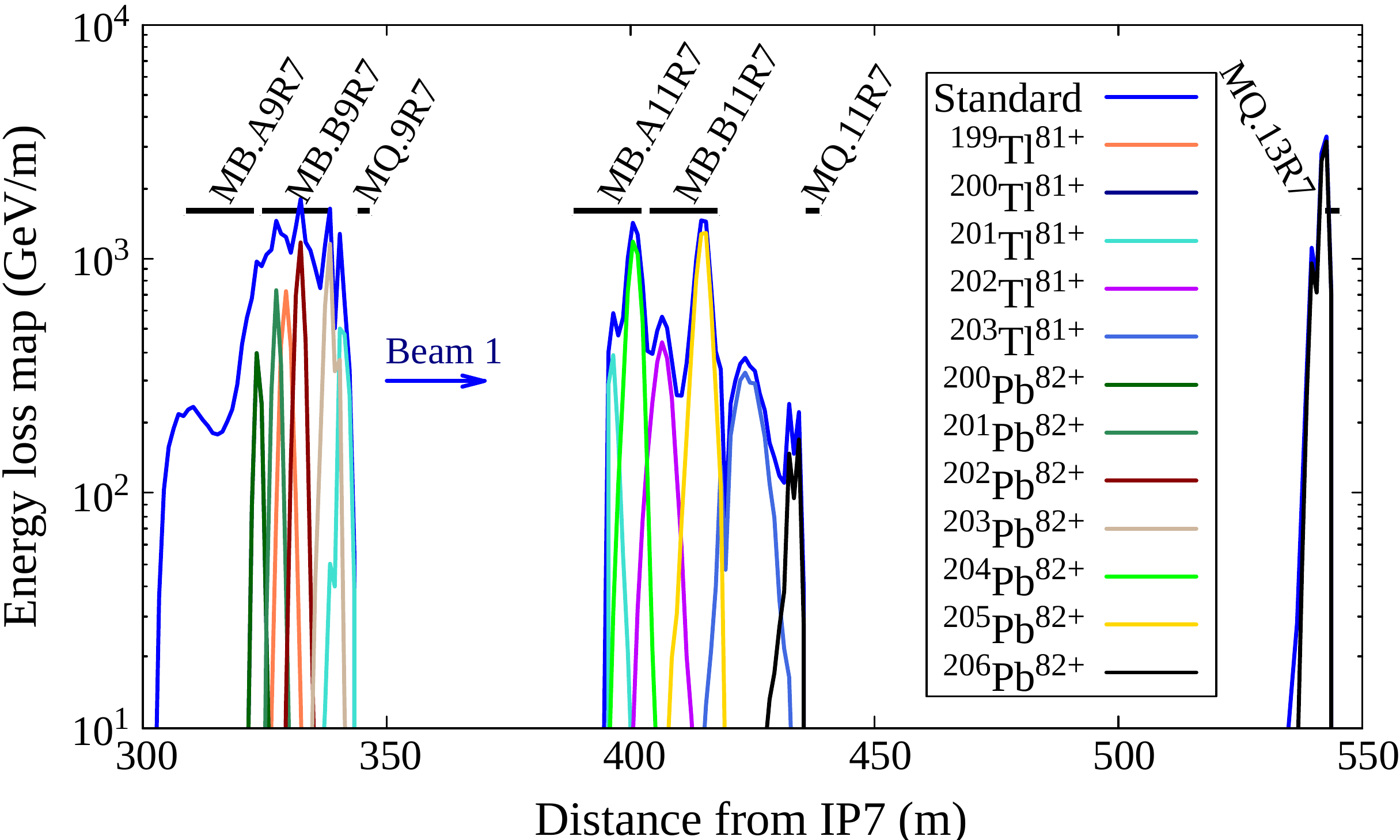}\\
\includegraphics[width=1\linewidth]{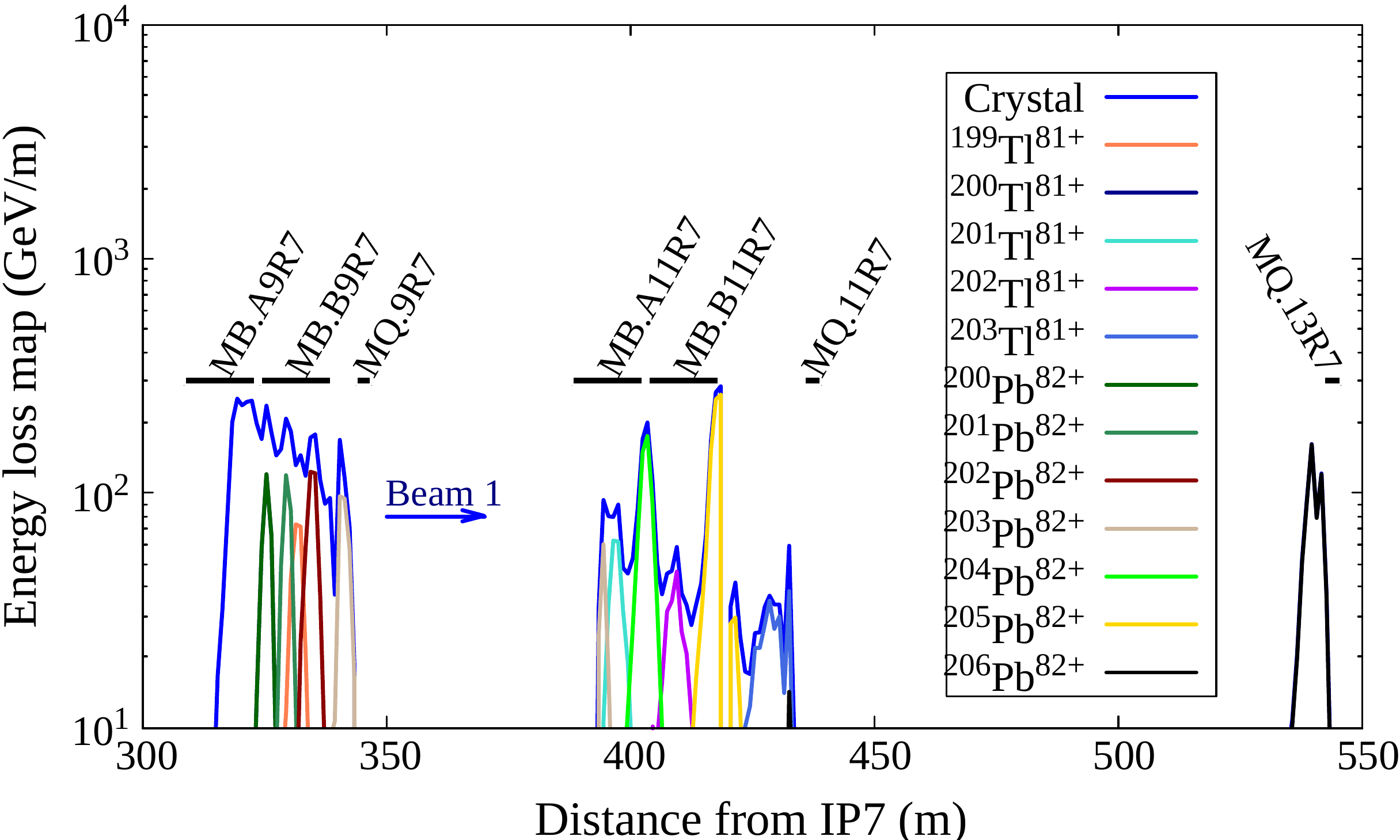}
\caption{Energy loss distribution on the cold magnet aperture downstream of the betatron cleaning insertion. The upper figure shows simulation results for the standard system and the lower figure for the crystal-assisted system. Results are given per $^{208}$Pb$^{82+}$ ion intercepted in the IR7 collimation system. Only fragments with kinetic energies $\geq$1~TeV/n were considered. The contribution of the most abundant heavy fragments are shown as separate curves. The simulation results correspond to the 2018 machine configuration (6.37~$Z$TeV).}
\label{fig:beam1_frag_coldaperture}
\end{figure}

Figure~\ref{fig:beam1_frag_coldaperture} shows the simulated energy loss distribution on the cold magnet aperture for the standard and crystal-assisted collimation systems, respectively.  The figure identifies the contributions of some of the most abundant heavy ion fragments with a magnetic rigidity allowing them to reach the dispersion suppressor and first arc cells. As discussed in Sec.~\ref{sec:benchmstdsystem}, the simulation for the standard system systematically underestimates the BLM signals at cold magnets. The BLM signals are closely correlated to the energy loss density on the cold magnet aperture, which is therefore expected to be underestimated by a similar factor as the BLM signals. In order to compensate for this difference, correction factors were applied on top of the simulation results in Fig.~\ref{fig:beam1_frag_coldaperture}, i.e., the results were scaled by the average ratio of measured and simulated BLM signals in each half-cell (factor of two for half-cells 9/11 and factor of five for half-cell 13). For the crystal-based setup, only the results in half-cell 13 were corrected, considering that the simulation overestimated the BLM signals in this cell (see Sec.\ref{sec:benchmcrysystem}).

In both collimation schemes, the energy loss distribution in half-cells~9, 11 and 13 is dominated by different lead and thallium isotopes, with a mix of lighter fragments (not shown) at the entrance of cell~9. The loss position is closely correlated to the magnetic rigidity and therefore the ion species. The created ion fragments have different charge-to-mass ratios and follow different dispersive orbits until they are lost; the dispersion suppressor dipoles therefore act as a spectrometer. Heavier isotopes of a given element are lost further downstream. Losses in cell 13 are composed of $^{206}$Pb$^{82+}$ ions, while lighter isotopes are lost in cells 9 and 11. Similar loss distributions were reported previously for the standard system \cite{hermes16_nim,Hermes_thesis}, yet the abundances in the different cells were different.
Considering the variety of fragments contributing to the energy loss distribution, the discrepancy between simulation and measurements in Fig.~\ref{Beam1_benchmark} is likely not due to individual isotope production yields in the primary collimator.  
This assumption is further underlined by the BLM benchmarks for cell~13, where the simulations respectively under- and overestimate the measurements for the standard and crystal-based collimation setup, respectively. 
In the standard collimation setup, hadronic interactions are responsible for producing 71\% of $^{204}$Pb$^{82+}$, 58\% of $^{205}$Pb$^{82+}$, and 49\% of $^{206}$Pb$^{82+}$ that are lost on the cold apertures, the rest is coming from EMD. On the other hand, in the crystal configuration, hadronic interactions contribute respectively to the production of 38\%, 29\% and 11\% of those fragments. 
Additionally, in the crystal setup, $^{205}$Pb$^{82+}$ and $^{204}$Pb$^{82+}$ are lost three meters further downstream compared to the standard collimation setup. 

\begin{figure}[!t]
\centering
\includegraphics[width=\linewidth,trim={0 1cm 2cm 2cm},clip,]{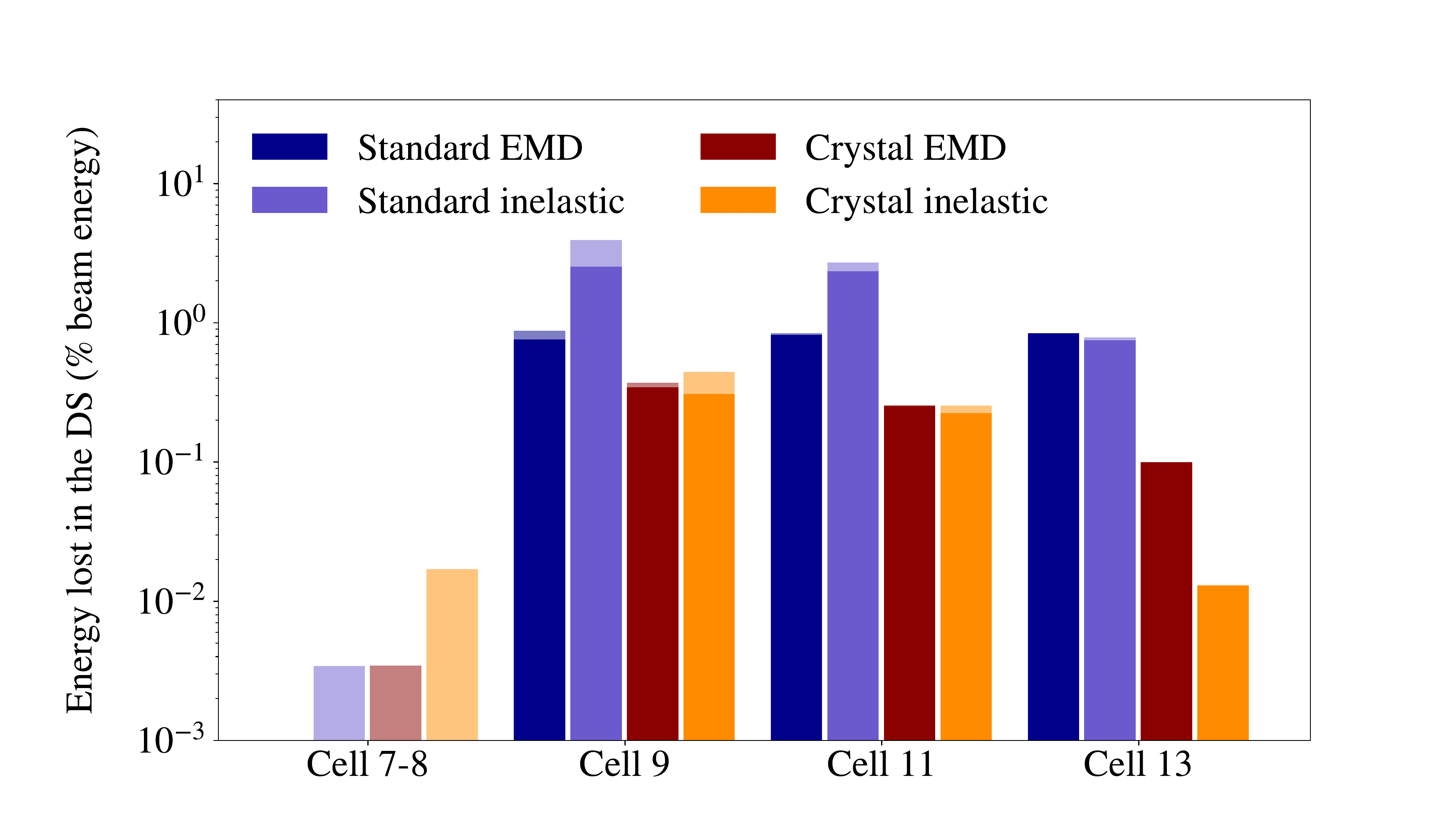}
\caption{Breakdown of the energy lost on the cold magnet aperture in different dispersion suppressor and arc half-cells. The figures indicates whether the original $^{208}$Pb$^{82+}$ ion was subject to hadronic or electromagnetic fragmentation in the primary collimator or crystal. Results are expressed in percentage of the total kinetic energy of all $^{208}$Pb$^{82+}$ ions intercepted in IR7. Darker colors correspond to the contribution of heavy fragments (Z$\geq$80).
The results correspond to the 2018 machine configuration (6.37~$Z$TeV).}
\label{fig:energy_lostdsarc}
\end{figure}

Figure~\ref{fig:energy_lostdsarc} quantifies the energy fraction lost in different half-cells of the DS and arc, showing separately the contribution of hadronic and electromagnetic $^{208}$Pb$^{82+}$ fragmentation products. The same empirical correction factors as above were applied on top of the simulations. The results are given per $^{208}$Pb$^{82+}$ ion intercepted in the IR7 collimation system. Emphasis is put on heavy fragments (Z~$\ge$~80 in darker colors) as they carry the main fraction of the power lost downstream of IR7 in both collimation setups. When normalized to the number of $^{208}$Pb$^{82+}$ collisions in the primary collimator or crystal, the average contribution of inelastic nuclear interactions to the energy loss in cold magnets is smaller than for EMD. This can be attributed to the different fragment distributions produced in the two types of interactions. Hadronic fragmentation products nevertheless dominate the overall energy leakage to cold magnets in the case of the standard system due to the much higher cross section compared to EMD. With crystals, the relative contribution of EMD and hadronic collision products becomes comparable in both half-cells 9 and 11, which are the most exposed cells in the cold region. 

\subsection{Fragment leakage with HL-LHC collimation settings}

The crystal-based collimation setup is planned to be used in regular $^{208}$Pb$^{82+}$ operation from 2023 (Run~3) onward. The beam intensity will be higher than in previous runs (see Table~\ref{tab:beamparam}), marking the start of the high-luminosity era for the heavy ion program. The collimator settings will be different than in the beam test studied above. In particular, the retraction of the primary collimators with respect to the crystal will be significantly smaller. This configuration provides a second protection layer. The nominal settings for future heavy ion runs are summarized in Table~\ref{tab:coll_settings_hllhc}. The crystal will be positioned 0.25$\sigma$ closer to the beam than the primary collimators, which are now located at 5$\sigma$. For completeness, the table repeats once more the 2018 settings used in the previous sections. 

In the following, we study the expected energy leakage to cold magnets with the future operational settings and compare the results to the standard system. For the standard system, we assume that both jaws of the horizontal primary collimator are placed at 5$\sigma$; compared to the 2018 settings (one jaw retracted by 0.5$\sigma$), this improves the system efficiency for DS losses and allows for a suitable comparison with the crystal-based system. As beam energy, we assume 7~$Z$TeV, which is the LHC design energy.
Furthermore, we consider the new collimator types with blocks made of MoGR (see also Sec.~\ref{sec:collimation}). The vertical and horizontal primary collimators, as well as selected secondary collimators, were already replaced with the new collimator type in the previous shutdown. Some secondary collimators will be replaced in the future. In this study, we assume that all but two secondary collimators are composed of Mo-coated MoGr blocks, as presently planned for Run~4. Like in the previous section, we consider beam halo losses in the horizontal plane, assuming an initial impact parameter of 1~$\mu$m.

\begin{table}[!t]
    \centering
    \begin{tabular}{lccc}
    \hline\hline
          & Regular system & \multicolumn{2}{c}{Crystal system}\\
         & (HL-LHC)  & (2018 test) & (HL-LHC)\\
        \hline
        TCPs & 5 & 9 & 5\\
        TCSGs$_{upstream}$&  6.5 & 8.6 & 6.5\\
        TCPCH &  / & 5 & 4.75\\
        TCSGs$_{downstream}$& 6.5 & 6.5 & 6.5\\
        TCLAs &  10 & 10 & 10\\
        \hline\hline
    \end{tabular}
    \caption{Collimator half gaps in IR7 during Run~2~\cite{NFuster2020} and HL-LHC heavy ion~\cite{DAndrea2019,DAndrea2021a}. Values are expressed in units of beam $\sigma$, calculated for a normalized proton beam emittance of 3.5~$\mu$m~rad.TCSGs$_{upstream}$ and TCSGs$_{downstream}$ indicate the TCSGs located on both sides of the horizontal crystal TCPCH.}
    \label{tab:coll_settings_hllhc}
\end{table}

\begin{figure}[!b]
\centering
\includegraphics[width=1\linewidth]{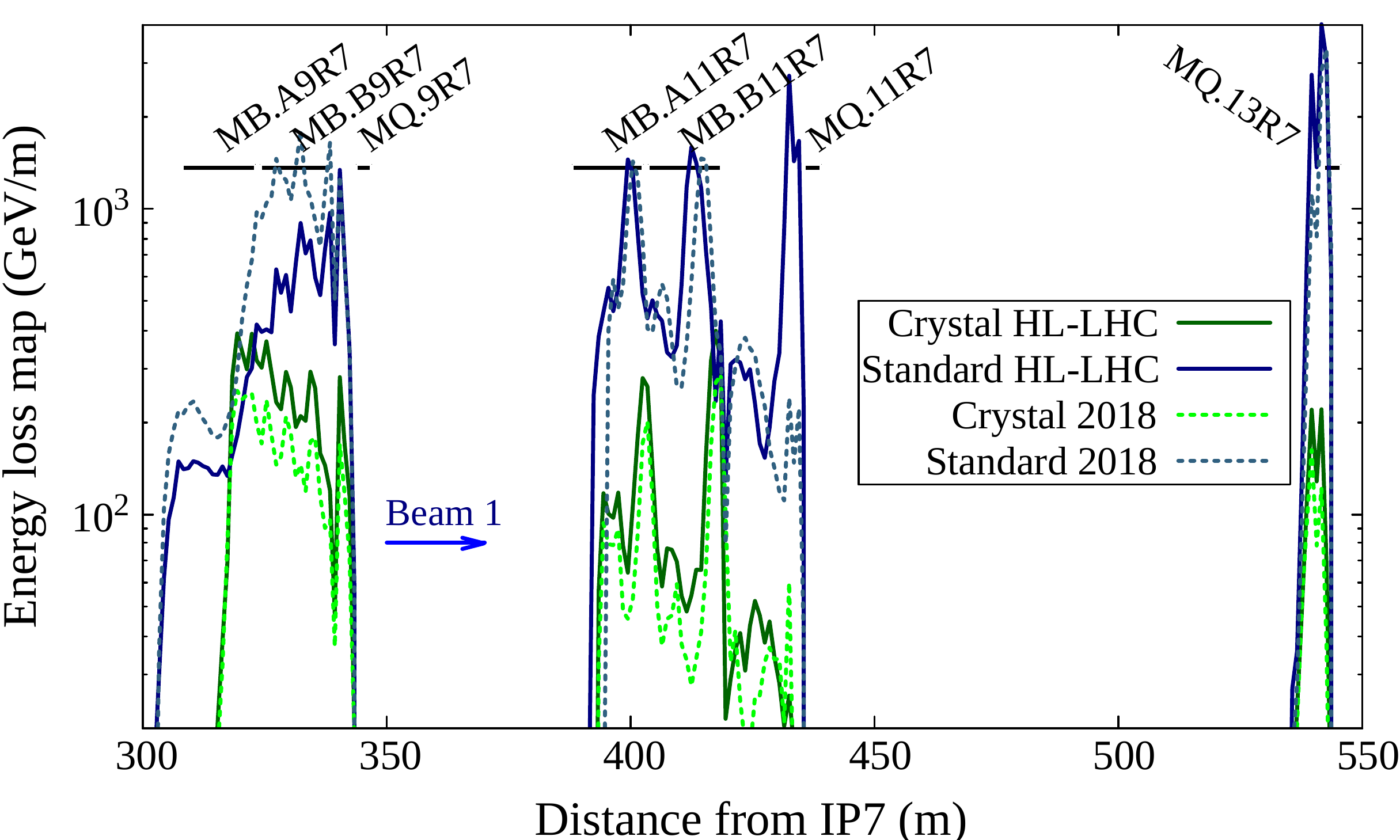}
\caption{Energy loss distribution on the cold magnet aperture downstream of the betatron cleaning insertion, comparing simulations for 2018 (dashed lines) and HL-LHC (solid lines) operational settings and beam parameters. Results are shown for the standard (in blue) and crystal-assisted (in green) collimation sytems, respectively. The curves are given per primary $^{208}$Pb$^{82+}$ ion intercepted in the IR7 collimation system during each simulation.}
\label{fig:energy_lossmap_HL}
\end{figure}
Figure~\ref{fig:energy_lossmap_HL} presents a comparison of the energy loss distribution on the cold magnet aperture for the standard and crystal-assisted collimation systems, with HL-LHC settings defined in Table~\ref{tab:coll_settings_hllhc} at 7~$Z$TeV and the 2018 machine configuration from Table~\ref{tab:coll_settings} at 6.37~$Z$TeV. The results include the same correction factors as in the previous section, accounting for the differences found in the benchmarks. Like above, the results are given per $^{208}$Pb$^{82+}$ ion intercepted in the IR7 collimation system. The two energy loss map for the crystal setups (in green) exhibit a similar pattern, but an increase of the energy loss density by up to 40\% can be observed with HL-LHC settings. A fraction of this increase can be attributed to the higher beam energy (7~$Z$TeV versus 6.37~$Z$TeV). 

The distributions for the standard system show some differences for 2018 and HL-LHC settings. This difference comes from the double-sided cleaning that was chosen as scenario to investigate a mock HL-LHC standard setup. It is known that the overall cleaning efficiency depends on the impacted jaw of the primary collimator~\cite{Hermes_thesis}. The simulations indicate that double-sided cleaning slightly decreases the quantity of fragments lost on the aperture of cell~9, however the left jaw of the primary collimator induces a spike of $^{206}$Pb$^{82+}$ losses at the end of cell~11. The different primary collimator materials (CFC versus MoGR) are found to have a marginal impact on the leakage to cold magnets. We will discuss this aspect in more detail in the next section, where the power deposition in magnet coils are presented.

\section{Power deposition in cold magnets with HL-LHC beams}
\label{sec:powdep_coldmagnets}

When secondary ions are lost on the aperture of cold magnets, they are subject to fragmentation and give rise to hadronic and electromagnetic showers. The shower development in the beam screen, cold bore and magnet components determines the power deposition density in the superconducting coils. This quantity, when compared to quench limits, is the primary indicator as to whether beam losses can lead to a quench. In this section, we present \textsc{FLUKA} power deposition simulations for the dispersion suppressor and arc cells downstream of IR7. The simulation accounts for the full particle shower development induced by the collision fragments described in the previous section. All results presented in this section are scaled to a beam lifetime of 0.2~hours, which is the design worth-case beam lifetime for HL-LHC heavy-ion runs. Assuming a beam intensity of 2.23$\times$10$^{11}$~$^{208}$Pb$^{82+}$ ions, this corresponds to a particle loss rate of 3.1$\times$10$^8$ $^{208}$Pb$^{82+}$ ions/s. All results assume the same HL-LHC collimator gaps as in the previous section (Table~\ref{tab:coll_settings_hllhc}), with a beam energy of 7~$Z$TeV. In the first part of this section, we present results for the standard collimation system, comparing the effect of different primary collimator materials (CFC and MoGR). In the second part, we compare the power deposition density for standard and crystal-based systems, respectively. In all cases, we assume that the ions impact on the horizontal primary collimator or crystal with an impact parameter of 1~$\mu$m. 

\subsection{Dependence on primary collimator material for the standard system}

\begin{figure}[!t]
\centering
\includegraphics[width=\linewidth]{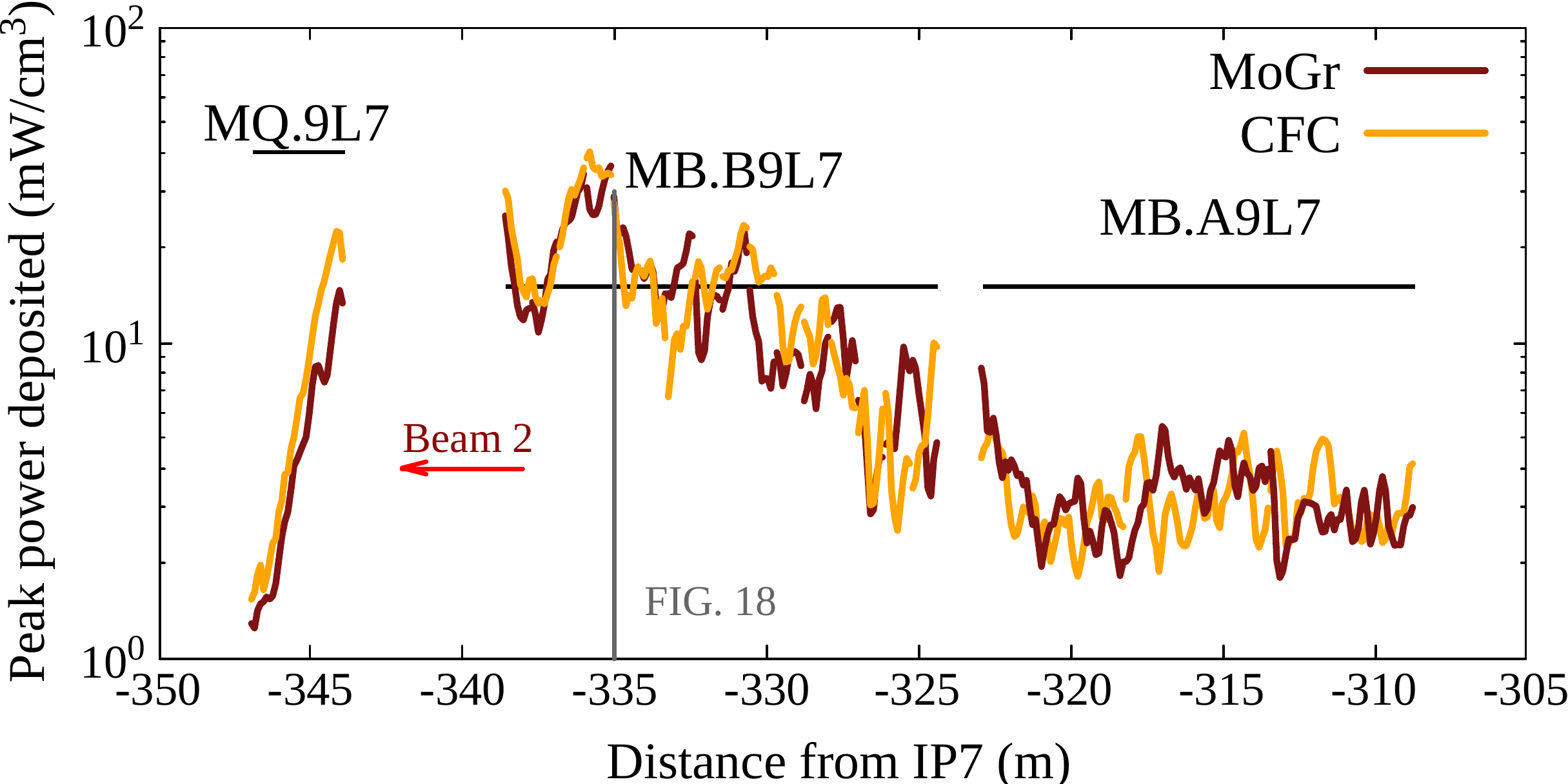}
\includegraphics[width=\linewidth]{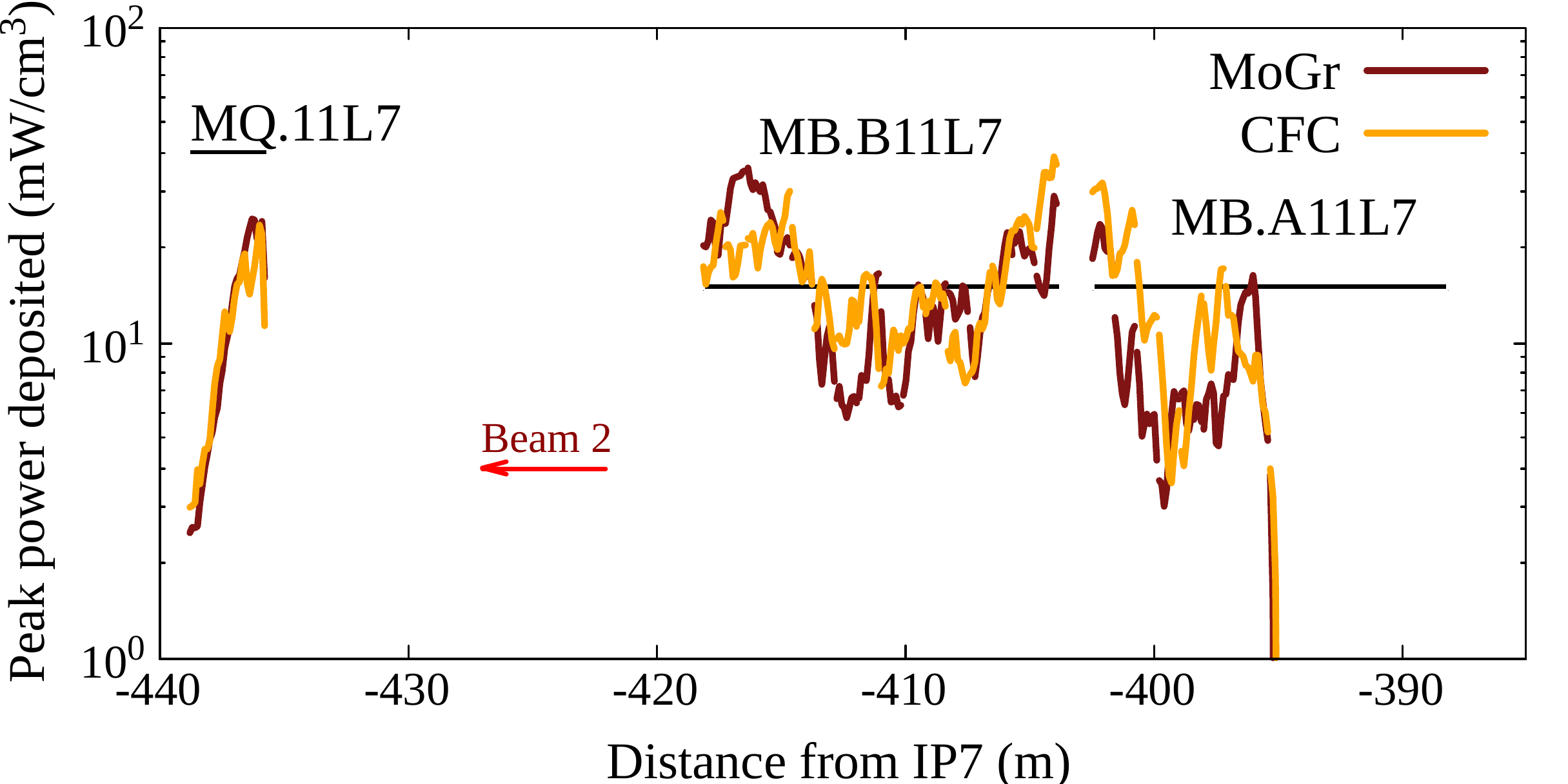}
\caption{Longitudinal distribution of the peak power density in the inner magnet coils for CFC and MoGR primary collimators, respectively. The figures show different half-cells of the dispersion suppressor downstream of IR7. The results correspond to betatron halo losses of 7~$Z$TeV $^{208}$Pb$^{82+}$ ions in the collimation system, assuming a beam lifetime of 0.2~hours. Black lines illustrate the magnet length and the respective estimated quench levels (MB - bending dipoles, MQ - lattice quadrupoles).}
\label{fig:energy_dep_mats}
\end{figure}

Figure~\ref{fig:energy_dep_mats} shows the peak power deposition density inside the superconducting magnet coils for CFC and MoGR primary collimators. Contrary to the previous section, the results correspond to the anticlockwise-rotating beam (beam 2). For consistency, we use the same correction factors as for beam 1 to compensate for the differences found in the benchmarks. During steady-state beam losses, the heat deposited by showers in the Rutherford cables has time to spread across the cable cross section~\cite{Auchmann2015}. The figure therefore shows radially averaged peak power density in the superconducting inner coils. The tentative quench levels are indicated by horizontal black lines; here we assume a quench level of 15~mW/cm$^3$ for dipoles at 7~$Z$TeV, considering that the quench level at this energy is lower than the 20~mW/cm$^3$ observed in the BFPP quench test at 6.37~$Z$TeV (see Sec.~\ref{sec:collimation}). For quadrupoles, we assume a quench level of about 40~mW/cm$^3$~\cite{Auchmann2015}. The statistical errors of the simulation are less than ten percent in the half-cells~9 and 11. 

As already found in previous studies for CFC collimators \cite{Aberle2020}, the simulation suggests that the quench levels of dipoles are exceeded in both cell~9 and cell~11. The figure also shows that the new material of the primary collimator (MoGR) does not significantly alter the power deposition density inside the superconducting magnet coils. As reported in Table~\ref{tab:physical_processes}, the mean free paths for hadronic fragmentation and EMD are different for CFC and MoGr. As a consequence, the relative fraction of $^{208}$Pb$^{82+}$ ions subject to EMD increases from about 14\% in CFC to about 25\% in MoGR. The different mean free paths also affect the reinteraction probability of secondary fragments inside the collimator blocks. These effects modify the composition of the secondary ion population. The change in the ion mass distribution is, however, not substantial enough to affect significantly the resulting power deposition in magnets. 

\begin{figure}[!t]
\centering
\includegraphics[width=0.8\linewidth]{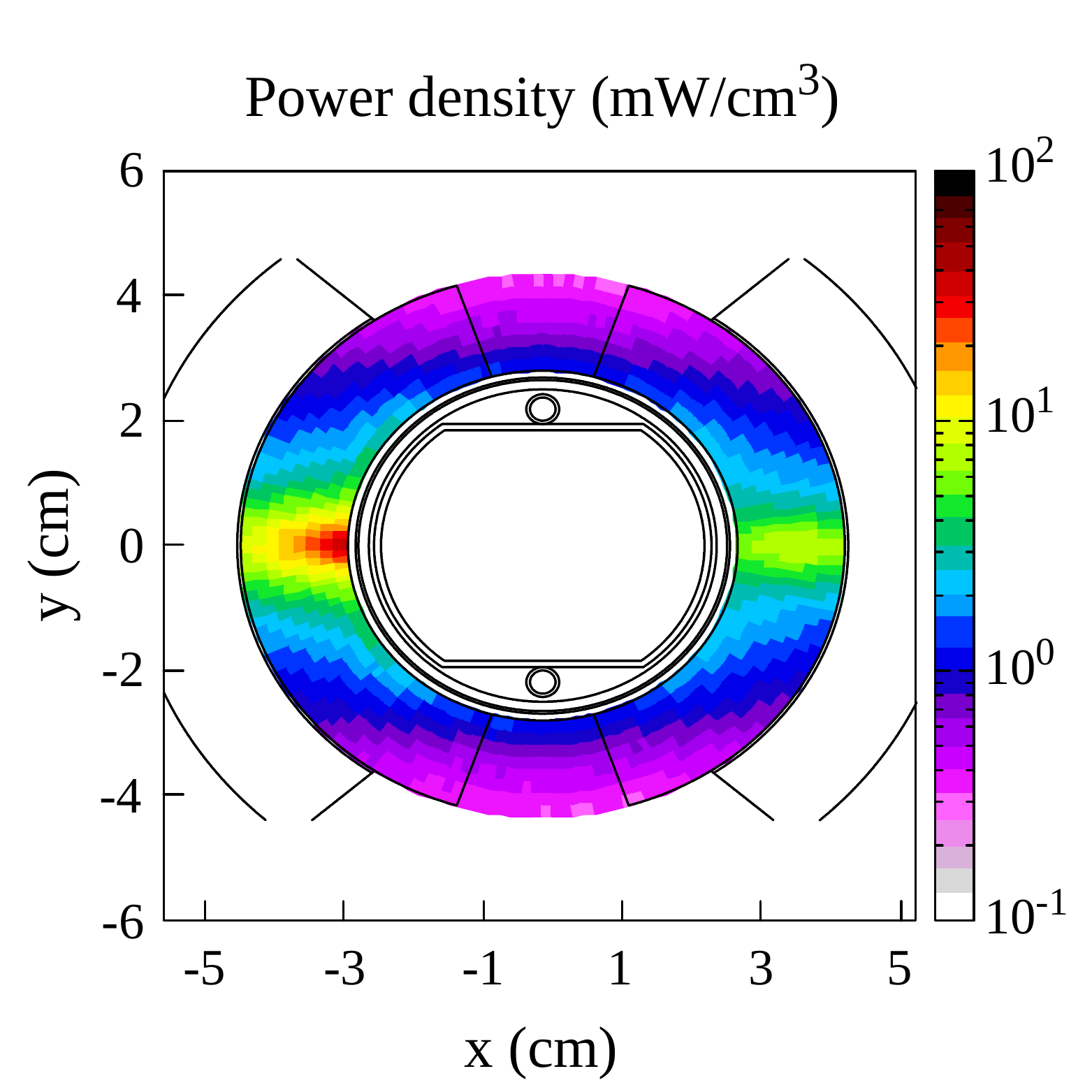}
\caption{Simulated transverse power density distribution in the most exposed dipole (MB.B9L7) in half-cell 9. The study corresponds to the standard collimation system with MoGR collimators (see also Fig.~\ref{fig:energy_dep_mats}). The beam direction enters the figure. The $x$-axis points towards the outer side of the LHC ring.}
\label{fig:energy_dep_rad}
\end{figure}

\begin{figure*}
\centering
\includegraphics[width=\linewidth,trim={0 0 0 0},clip]{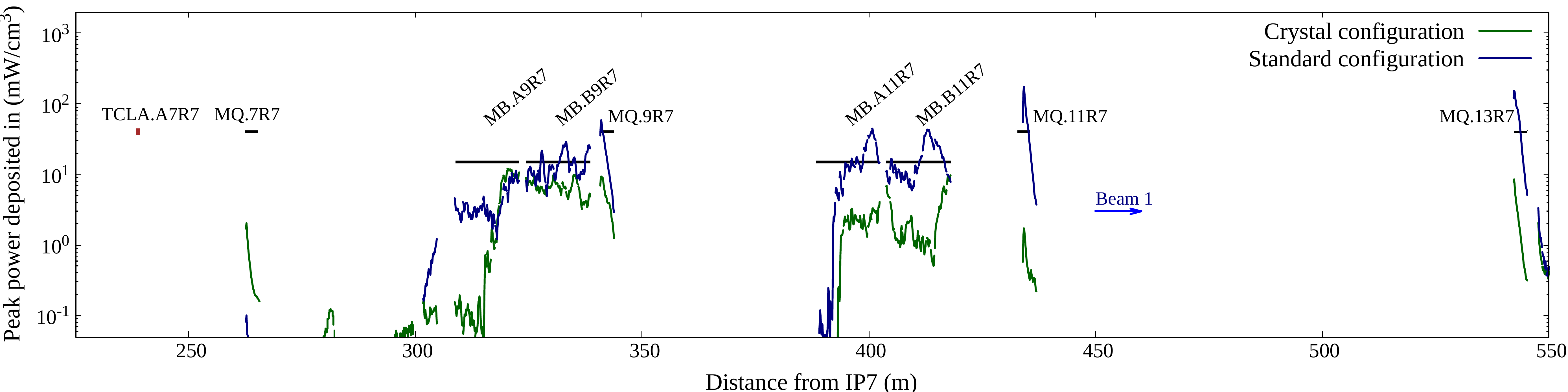}
\caption{Longitudinal distribution of the peak power density in the inner magnet coils for standard and crystal-assisted collimation, respectively. The results correspond to betatron halo losses of 7~$Z$TeV $^{208}$Pb$^{82+}$ ions in the collimation system, assuming a beam lifetime of 0.2~hours. Black lines illustrate the magnet length and the respective estimated quench levels (MB - bending dipoles, MQ - lattice quadrupoles)}
\label{fig:Energy_HL}
\end{figure*}
Figure~\ref{fig:energy_dep_rad} shows the transverse power density map in the inner coil of the most exposed dipole (MB.B9L7 in Fig.~\ref{fig:energy_dep_mats}). The map corresponds to the longitudinal position, where the power density is maximum as indicated in Fig.~\ref{fig:Energy_HL} by the vertical grey line. Most of the secondary ion fragments are overbent by the dipole field and therefore impact on the inner side of the magnet aperture. This creates a local hot spot on the magnet midplane as indicated in the figure. The most notable exception of secondary fragments lost on the outer side of the aperture are $^{3}$H$^{1+}$ ions, which have a lower charge-to-mass ratio than the beam particles. The $^{3}$H$^{1+}$ ions carry, however, only a small fraction of the power lost in cold magnets.   

\subsection{Standard versus crystal-assisted collimation}

Figure~\ref{fig:Energy_HL} presents the longitudinal peak power density profile in the coils of DS and arc magnets for the standard and the crystal-assisted collimation system, respectively. The results are for the clockwise-rotating beam (beam~1). Like in the previous section, the power densities were radially averaged over the cable width and were corrected empirically based on the differences found in the simulation benchmarks. 

As can be seen in the figure, the crystal reduces the power deposition density in most of the magnets. The only notable exception is the quadrupole in half-cell~7 (MQ.07R7 in Fig.~\ref{fig:Energy_HL}), where a twentyfold increase of the power density can be observed. This increase was also seen in the beam tests in 2018 \cite{DAndrea2019,DAndrea2021,Dandrea_2023} and is mainly due to a higher number of heavy-ion fragments ($Z>65$) impacting the last TCLA collimator. This collimator is located about 24~meters upstream of the quadrupole. Being made of a tungsten-alloy, the TCLA has a good absorbing capability, but secondary shower particles (mainly protons and neutrons) can still reach the Q7. Nevertheless, the power deposition density in this magnet remains around a factor of twenty below the expected quench level. The increased losses on the TCLA are therefore not expected to pose a bottleneck for crystal-assisted collimation.

The maximum power density in the first dipole of half-cell~9 (MB.A9R7) is very similar for both systems. Nevertheless, the crystal reduces the power density in the second, more exposed dipole (MB.B9R7). With the crystal, the maximum power density in the dipoles is 12~mW/cm$^3$, which is below the expected quench level. The relative gain with the crystal setup is more pronounced in half-cell 11; the maximum power density in the dipoles is reduced by at least a factor of four. A slight shift of longitudinal power density distribution can be observed between the two setups. With the crystal, the two peaks induced by $^{204}$Pb$^{82+}$ and $^{205}$Pb$^{204}$ ions occur both in the second dipole of half-cell 11; with the standard system, the $^{204}$Pb$^{82+}$-induced peak occurs in the first dipole. The simulation suggests that the crystal setup can reduce the power density in the two dipoles below the expected quench level; more precisely, in the second more exposed dipole (MB.B11R7), the maximum obtained power density is less than 10~mW/cm$^3$. The simulation also indicates that the crystal mitigates the possible risk of quenches in the three lattice quadrupoles in the half-cells~9, 11 and 13.

Some uncertainty still remains about the actual quench margin with the crystal-based collimation system. Since the heavy-ion run in 2018, new crystals have been installed in the machine.
Furthermore, some uncertainty remains concerning the actual quench level of dipoles at 7~$Z$TeV as the quench margin at this energy could not yet be validated experimentally. 
The results obtained in this section nevertheless suggest that crystal-assisted collimation has the potential to avoid magnet quenches even in case of lifetime drops to 0.2~hours. 

Similar power deposition studies have been carried out previously for the option of a dispersion suppressor collimator with new 11~T magnets. As indicated in the introduction, the installation of this setup has been deferred. For this configuration, the simulations indicated a factor of two quench margin for the most exposed magnet (11~T dipole) \cite{Waets2021}, since the 11~T dipoles are expected to have a higher quench level than standard bending dipoles. The present studies suggests that a crystal-based setup cannot yield the same margin, but is still a very promising baseline for Run~3.

\section{Conclusions}
\label{sec:conclusions}

The betatron halo collimation system is a key system for assuring the operational performance of the LHC in both proton and heavy-ion runs. The LHC heavy-ion program with $^{208}$Pb$^{82+}$ beams will  benefit from a significant increase of the beam intensity in Run~3. The stored beam energy is expected to surpass 20~MJ. Despite the excellent performance of the collimation system in past runs, secondary ion fragments from collimators may pose a performance limitation in future high-intensity $^{208}$Pb$^{82+}$ operation since they can lead to magnet quenches in case of beam lifetime drops, which cannot be  excluded at high beam current operation. For this reason, it has been foreseen to use bent crystals as primary collimators in order to mitigate the risk of quenches in absence of the baseline dispersion suppressor collimators. In this paper, we presented first power deposition studies for cold magnets, which provided an absolute comparison between the two collimation techniques. The calculations were based on a previously developed and well benchmarked simulation chain, consisting of particle tracking and beam-matter interaction studies.

The paper presented absolute benchmarks of the simulation chain against BLM measurements from the 2018 $^{208}$Pb$^{82+}$ run at 6.37~$Z$TeV. A series of fast beam loss events was observed in 2018, which provided a valuable reference for benchmarking the model calculations for the standard collimation system. Taking into account the complexity of ion-matter interactions and fragment production in collimators, the simulation chain gave a satisfying agreement with measurements. In particular, the BLM signals in the dispersion suppressor were reproduced within a factor of two. This agreement is better than was found in a previous benchmark for the counter-rotating beam, which can mainly be attributed to the different initial conditions. We also presented a first absolute BLM simulation benchmark for the crystal-based setup. Crystals were not used in regular heavy-ion physics operation in past runs, but their performance was evaluated in dedicated beam tests. Based on the BLM measurements recorded during one of these tests, the predictive ability of the crystal simulation setup could be probed. The simulation relied on multi-turn tracking simulations with all ring collimators and a dedicated \textsc{FLUKA} model for describing coherent effects in strip crystals. A very good agreement was found for the dispersion suppressor region, confirming the suitability of the model for making predictions for future runs.    

The simulations show that a large fraction of the power deposited in the most exposed cold magnets is due to secondary fragments produced in the primary collimator or the crystal. The studies also indicated that the fraction of collisions contributing to the leakage is significantly less for primary ions that received a kick due to channeling in crystals. The results therefore suggest that the efficiency of the crystal-based system is rather independent of the actual suppression of hadronic or electromagnetic collisions in channeling. The main advantage of the crystal is the reduced number of $^{208}$Pb$^{82+}$ fragmentation processes in the amorphous regime, since a large fraction of the ions is directed onto a secondary collimator. Fragmentation of ions in this secondary collimator yields a very small contribution to the particle leakage to cold magnets, due to their large impact parameter.

In the last part of the paper, we provided predictions of the power deposition density inside superconducting coils, with the goal to estimate the quench margin with crystal-assisted collimation in future heavy-ion runs. We simulated a scenario, where the beam lifetime drops to 0.2~hours, which is the specified design loss scenario that the collimation system should be able to handle without a beam dump or a magnet quench. The simulations were corrected for the differences found in the benchmarks. The studies indicate the crystal-based collimation setup can provide a satisfactory reduction of the power density in superconducting magnets located downstream of the betatron cleaning insertion. Assuming HL-LHC beam parameters and the empirical corrections, the power deposition density is predicted to remain below the expected quench levels for all the magnet families. The margin might, however, be only a few tens of percent, depending on the actual quench level of dipoles and on the actual simulation error margin. The studies presented in this paper focused on one specific scenario (losses in the horizontal plane of the clockwise-rotating beam). A separate assessment is needed for the other plane and the other beam, following the procedure outlined in this paper, although the results are expected to be qualitatively similar. Particles lost on the cold aperture from the vertical plane are less critical.

\section*{Acknowledgements}

This work was supported by the High Luminosity Large Hadron Collider project.

\bibliography{apssamp}

\end{document}